\documentclass[useAMS,usenatbib]{mn2e}

\usepackage{epsfig}

\usepackage{color}
\usepackage{amsmath}    
\usepackage{mathtools}  
\usepackage{amssymb}    
\usepackage{verbatim}   
\usepackage{hyperref}   
\usepackage{multicol}  %
\usepackage{multirow}
\usepackage{ifthen}  

\usepackage{graphicx}
\usepackage{caption}
\usepackage{subcaption}

\usepackage{lipsum}       
\usepackage{changepage}   
\usepackage{enumitem}
\usepackage{multirow}

\usepackage{threeparttable}

\newcommand{\hinv}{\ensuremath{\, h^{-1}}}%
\newcommand{\msol}{\ensuremath{\, {\rm M}_\odot}}    
\newcommand{\msun}{\ensuremath{\, {\rm M}_\odot}} 
\newcommand{\kpc}{\ensuremath{\, {\rm kpc}}}         
\newcommand{\mpc}{\ensuremath{\, {\rm Mpc}}}         
\newcommand{\gpc}{\ensuremath{\, {\rm Gpc}}}%

\newcommand{\apj}{ApJ}

\newcommand{\aap}{A{\&}A}

\newcommand{\mnras}{MNRAS}

\newcommand{\araa}{ARAA}

\newcommand{\physrep}{PhysRep}
\newcommand{\prd}{Phys. Rev. D}

\newcommand{\mgas}{{\rm M}_{\rm gas}}   
\newcommand{\mstar}{{\rm M}_{\rm star}}
\newcommand{\mhalo}{{\rm M}_{\rm halo}}

\newcommand{\MFmu}{\frac{dn(\mu,z)}{d\mu}}
\newcommand{\varmassfullone}{\sigma_{\mu | s}^2}

\newcommand{\bahamas}{\textsc{bahamas} } 
\newcommand{\macsis}{\textsc{macsis} }

\defcitealias{Evrard:2014}{E14}

\title[Massive halo property covariance] {Localized Massive halo properties in BAHAMAS and MACSIS simulations: scalings, log-normality, and covariance}

\author[A.~Farahi et al.]
{Arya Farahi$^{1}$\thanks{E-mail: \texttt{aryaf@umich.edu}},
August E. Evrard$^{1,2}$,
Ian McCarthy$^{3}$,
David J. Barnes$^{4,5}$, and
\newauthor
Scott T. Kay$^{4}$
\vspace{2mm}\\
	$^1$Department of Physics and Michigan Center for Theoretical Physics, University of Michigan, Ann Arbor, MI 48109, USA\\
    $^2$Department of Astronomy, University of Michigan, Ann Arbor, MI 48109, USA\\
    $^3$Astrophysics Research Institute, Liverpool John Moores University, 146 Brownlow Hill, Liverpool L3 5RF\\
    $^4$Jodrell Bank Centre for Astrophysics, School of Physics and Astronomy, The University of Manchester, Manchester M13 9PL, UK\\
    $^5$Department of Physics, Kavli Institute for Astrophysics and Space Research, Massachusetts Institute of Technology, Cambridge, MA 02139, USA
}

\begin{document}

\date{Accepted. Received; in original form}

\pagerange{\pageref{firstpage}--\pageref{lastpage}} \pubyear{2014}

\maketitle

\label{firstpage}

\begin{abstract}
Using tens of thousands of halos realized in the \bahamas and \macsis simulations produced with a consistent astrophysics treatment that includes AGN feedback, we validate a multi-property statistical model for the stellar and hot gas mass behavior in halos hosting groups and clusters of galaxies. 
The large sample size allows us to extract fine-scale mass--property relations (MPRs) by performing local linear regression (LLR) on individual halo stellar mass ($\mstar$) and hot gas mass ($\mgas$) as a function of total halo mass ($\mhalo$). We find that: 1) both the local slope and variance of the MPRs run with mass (primarily) and redshift (secondarily); 2) the conditional likelihood, $p(\mstar,\ \mgas  | \ \mhalo, z)$ is accurately described by a multivariate, log-normal distribution, and; 3) the covariance of $\mstar$ and $\mgas$ at fixed $\mhalo$ is generally negative, reflecting a partially closed baryon box model for high mass halos.  We validate the analytical population model of \citet{Evrard:2014}, finding sub-percent accuracy in the log-mean halo mass selected at fixed property, $\langle \ln \mhalo | \mgas \rangle$ or  $\langle \ln \mhalo | \mstar \rangle$, when scale-dependent MPR parameters are employed. This work highlights the potential importance of allowing for running in the slope and scatter of MPRs when modeling cluster counts for cosmological studies. We tabulate LLR fit parameters as a function of halo mass at $z=0$, $0.5$ and 1 for two popular mass conventions.
\end{abstract}


\section{Introduction}

Dark matter halos provide the gravitational potential wells within which baryonic plasma can cool and form stars and galaxies.  Measuring galaxy assembly across cosmic history is key to understanding the astrophysical processes happening within halos.  Over the past two decades it has become clear that the highest mass halos that host groups and clusters of galaxies are, in an overall sense, less efficient at converting baryons into stars.   The majority of baryons end up in a hot intracluster medium (ICM) \citep{Briel:1992}. Despite the inefficiency of star formation within the overall halo, the central galaxies of groups and clusters are the largest in the universe, built by merging and accretion of many smaller systems \citep[{\sl e.g.},][]{Richstone:1976,DeLucia:2007}.

Considerable effort has gone into measuring the statistical relationship between the mass and observable properties of halos that reflect their baryon contents \citep[see][ for a recent review]{Giodini:2013}.  Observational studies are limited by sample of tens to low hundreds, systematic uncertainties in total mass estimates, and complex or ill-defined sample selection criteria.  Recent efforts are improving on these fronts \citep{Mantz:2016,Mantz:2016b,Zou:2016,Saro:2017,Schellenberger:2017}. 

We use the term mass--property relation (MPR) to represent the functional form of conditional halo statistics, $p(\mathbf{S} | M,z)$, where $\mathbf{S}$ is a set of intrinsic properties of the population of halos of mass $M$ at redshift $z$. We use the term {\sl property} rather than {\sl observable} here intentionally, as our work involves three-dimensional spatial measurements of stellar and hot gas mass properties at specific radii in simulations.  While not directly observable, estimators for these quantities can be constructed from optical, X-ray or SZ observations. 
Knowledge of the MPR, and survey-specific mappings to observed quantities, are critical for understanding multi-phase baryon evolution and for producing competitive cosmological constraints using cluster counts \citep{Allen:2011, Weinberg:2013}.  

Cosmological hydrodynamical simulations that evolve gravitationally-coupled baryons and dark matter offer model-dependent predictions for the form and redshift evolution of massive halo scaling relations \citep[{\sl e.g.},][]{Evrard:1996,BryanNorman:1998,Sembolini:2013,LeBrun:2016,Barnes:2017}.  While significant progress has been made, multi-fluid hydrodynamic simulations remain challenged by the wide dynamic range and complex astrophysical elements involved in modeling the formation of stars, supernova feedback, and supermassive black hole effects.
The \bahamas simulations \citep{McCarthy:2017} have taken a novel approach by tuning sub-grid control parameters to match the observed galaxy stellar mass function and the hot gas mass fractions of groups and clusters simultaneously.  
The \bahamas simulations are a set of $400 \mpc \hinv$ volumes that includes metal-dependent radiative cooling, star formation, and prescriptions for both supernova and AGN feedback. 
This suite of simulations reproduce a wide range of observables and have been used to characterize biases in a broad range of mass estimation techniques \citep{Henson:2017}. 

Multi-wavelength population statistics require understanding the covariance between pairs of intrinsic properties or observable quantities.  This covariance is an essential element in modeling multi-wavelength cluster samples, as pointed out by \citet{Nord:2008} for the case of inferring luminosity evolution from X-ray flux-limited samples.

The diagonal elements of the covariance matrix linking mass to observable properties are becoming better measured, but currently off-diagonal elements are poorly known \citep{Mantz:2016}. 
Cosmological hydrodynamics simulations, however, are a great tool for gaining insight into the detailed form of the MPR, including property covariance.

The likelihood of little or no loss of baryons from the deepest potential wells motivates an expectation of anti-correlation in the gas and stellar mass fractions in the highest massive halos. If all clusters of fixed halo mass are closed baryon boxes with baryons partitioned into stars and gas, then a particular system with slightly more (less) gas than average must contain a lower (higher) stellar mass than average, meaning a strong anti-correlation between gas mass and stellar mass.
Such an anti-correlation is apparent in the Rhapsody-G simulations of \citet{Wu:2015}, where a correlation coefficient $r = -0.7$ is found for gas and stellar mass deviations about the mean in a sample of ten $10^{15} \msol$ halos and their progenitors. 

In lower-mass halos hosting groups and poor clusters of galaxies, feedback can effectively drive baryons outside of the virial radius \citep[e.g. ][]{Lau:2010,Sembolini:2013,Truong:2016,LeBrun:2016}, 
reducing or eliminating the degree of anti-correlation.
 
Another key assumption in modeling MPRs 
is the form of the conditional distribution of properties at fixed halo mass, usually assumed to take a log-normal form.  
Under a log-normal assumption coupled with a simple parameterized approximation to the halo space density, or mass function, \citet[][hereafter E14]{Evrard:2014} derive closed-form expressions for multi-property population statistics.  
The analytic model exposes fundamental parameter degeneracies between the shape of the mass function, which is driven by cosmology, and MPR parameters determined by astrophysical processes.  Practically, the model supports fast computation of expectations for cosmological likelihood analysis.

The goals of this work are: i) to measure the mass and redshift dependencies of MPRs for stellar mass and hot gas mass; ii) evaluate the statistical form of the MPR likelihood, and; iii) test the accuracy of the \citetalias{Evrard:2014} model in a simulation setting where the intrinsic properties are measured directly.  
Unlike previous ``zoom-in'' simulations \citep[{\sl e.g.},][]{Wu:2015}, the \bahamas simulation models baryon behavior in a large cosmic volume, enabling study of a wide range of halos hosting groups and clusters. The large samples from \bahamas allow us to apply a localized regression approach to estimate mass-dependent MPR parameters.  However, the $400 \hinv \mpc$ simulation size limits the number of the most massive halos; \bahamas statistical coverage drops off above $3 \times 10^{14} \msol$.  We therefore also include the \macsis simulation ensemble which, like \citet{Wu:2015}, uses the zoom-in technique to extend the mass range of the \bahamas sample while employing the same astrophysical model, resolution, and cosmology \citep{Barnes:2017}.

This paper organized as follows. In \S\ref{sec:simulation} we present the simulation samples used in this work while \S\ref{sec:model} describes our non-parametric local linear regression (LLR) model. The LLR results, including covariance of hot gas and stellar mass at fixed halo mass, are presented in \S\ref{sec:results}.  In \S\ref{sec:E14-test} we test the performance of the \citetalias{Evrard:2014} analytic model, followed by discussion in \S\ref{sec:discussion} and a summary in \S\ref{sec:conclusion}.

Throughout this paper, we use radial and mass scales defined by a spherical density contrast with respect to the critical density of the
universe, $\rho_{\rm crit}(z)$; ${\rm M}_{\Delta}$ indicates the mass within which the average total mass density is $\Delta \rho_{\rm crit}(z)$.  Halo masses are expressed in units of $\msun$,  not $\hinv \msun$).

\section{Simulations}\label{sec:simulation}

We use the \bahamas cosmological hydrodynamical simulation \citep{McCarthy:2017} run using the Gadget-3 SPH code with subgrid prescriptions for metal-dependent radiative cooling, star formation, and stellar and AGN feedback developed as part of the OverWhelmingly Large Simulations project \citep{Schaye:2010}.  The periodic $400 \hinv \mpc$ cube we use here adopts a flat $\Lambda$CDM cosmology with \emph{Planck 2013} cosmological parameters \citep{Planck:2014}, namely ${\Omega_{m}, \Omega_{b}, \Omega_{\Lambda}, \sigma_{8}, n_{s}, h} = {0.3175, 0.049, 0.6825, 0.834, 0.9624, 0.6711}$ where $\Omega_m$, $\Omega_b$ and $\Omega_\Lambda$ are the normalized densities in matter, baryons and vacuum energy, $\sigma_8$ sets the power spectrum normalization, $n_s$ is the primordial spectral index, and $h \equiv H_0/(100~{\rm km~s}^{-1}~{\rm Mpc}^{-1})$ is the dimensionless Hubble constant.

The wind velocity associated with stellar feedback and the heating temperature associated with the AGN feedback in \bahamas are adjusted so as to reproduce the observed local galaxy stellar mass function and the amplitude of the relation between hot gas mass and halo mass of local X-ray-selected galaxy groups and clusters.  Non-tuned features match an unprecedentedly wide range of observed properties, including galaxy and hot gas radial profiles as well as the behavior of stacked SZ and X-ray luminosity as a function of galaxy stellar mass \citep{McCarthy:2017}.

Cosmological simulations featuring volume-complete hydrodynamics with full sub-grid physics at high spatial and mass resolution are very computationally expensive.  
The $400 \hinv \mpc$ \bahamas simulation has spatial resolution of $4 \hinv \kpc$ and resolves a $10^{14} \msol$ halo with $\sim 30,000$ particles.  Because of the limited number of very high mass halos in the realized volume, the \macsis project \citep{Barnes:2017} was developed to extend the sample to higher mass halos. The \macsis ensemble consists of 390 ``zoom-in'' simulations \citep{Tormen:1997} of individual halo regions drawn from a parent $3.2 \gpc$ N-body simulation.  The hydrodynamic resimulations employ the same resolution and sub-grid prescriptions as \bahamas in a Planck cosmology with nearly identical parameters as \bahamas  \citep[parameter values typically differ in the third significant digit, see][]{Barnes:2017}.  

As described in \citet{McCarthy:2017}, halos are identified using a ``friends-of-friends'' percolation method.  The spherically integrated quantities used here are measured using the minimum of the local gravitational potential as the halo center, and any sub-halos that lie outside the characteristic radii, $R_{\Delta}$ are ignored. 

\begin{table}
\centering
\caption{Halo sample sizes with $M_{500}>10^{13} \msol$.}
\label{tab:NumHalos}
\begin{tabular}{|c|c|c|}
\hline
Redshift & \bahamas & \macsis \\ 
\hline
1  & 11387  & 377 \\
0.5/0.46$^a$ & 17668  & 377 \\
0 & 21987  & 385 \\
\hline
$^a$ 0.5=\bahamas, 0.46=\macsis 
\end{tabular}
\end{table}

The samples we use, listed in Table~\ref{tab:NumHalos}, include all halos with $M_{500} > 10^{13} \msun$ at redshifts $z=0$, $0.5$ and $1.0$. Note that there the redshift slice for \macsis sample is $0.46$. The combined \bahamas and \macsis simulations offer tens of thousands of halo realizations covering a wide dynamic range in total mass.  

The halo properties we study are the aggregate stellar mass, $\mstar$, and the hot phase gas mass, $\mgas$, measured within spheres enclosing densities of $\Delta=500$ and $200$ times the critical density, $\rho_{\rm crit}(z)$.  
 Note that the hot gas mass includes particles with temperatures greater than $10^{5}$~K while the stellar mass uses all star particles within $R_{\Delta}$.

For this study, we combine \bahamas and \macsis samples into a super-sample. Since the \bahamas and \macsis are not using exactly the same cosmology, we re-normalize the baryonic contents of the \macsis sample to align the global baryon fraction, $\omega_b / \Omega_m$, to that assumed in the \bahamas cosmology; however, the magnitude of this correction is negligible, $<2\%$.  We also note that there is small difference in the redshift of \bahamas and \macsis samples, 0.5 versus 0.46.  Since we show below that the redshift evolution of the properties we examine is relatively weak, we do not apply any correction for this redshift. 

The complex interactions of mergers, turbulence, cooling, chemical enrichment, and feedback from supernovae and AGN play out within the evolving cosmic web network of large-scale structure to determine the overall statistical nature of the baryon component masses within the halo population.  While matching observed mean stellar and gas fraction behavior, within the limits of current observational uncertainties, has been done in the \bahamas and \macsis simulations by tuning a small number of sub-grid parameters, higher-order features of the property statistics should be considered {\sl model-dependent predictions} of the underlying astrophysical theory.  Within the context of these simulations' numerical and astrophysical treatments, we focus this paper on the model's expectations for running of the slope and scatter of the MPR with mass and redshift.  Future work can examine the robustness of these features using multiple simulations by independent groups.

\section{Mass-localized Regression}\label{sec:model}

\begin{figure*}
  \begin{subfigure}[b]{0.48\textwidth}
       \centering 
    	\includegraphics[width=\textwidth]{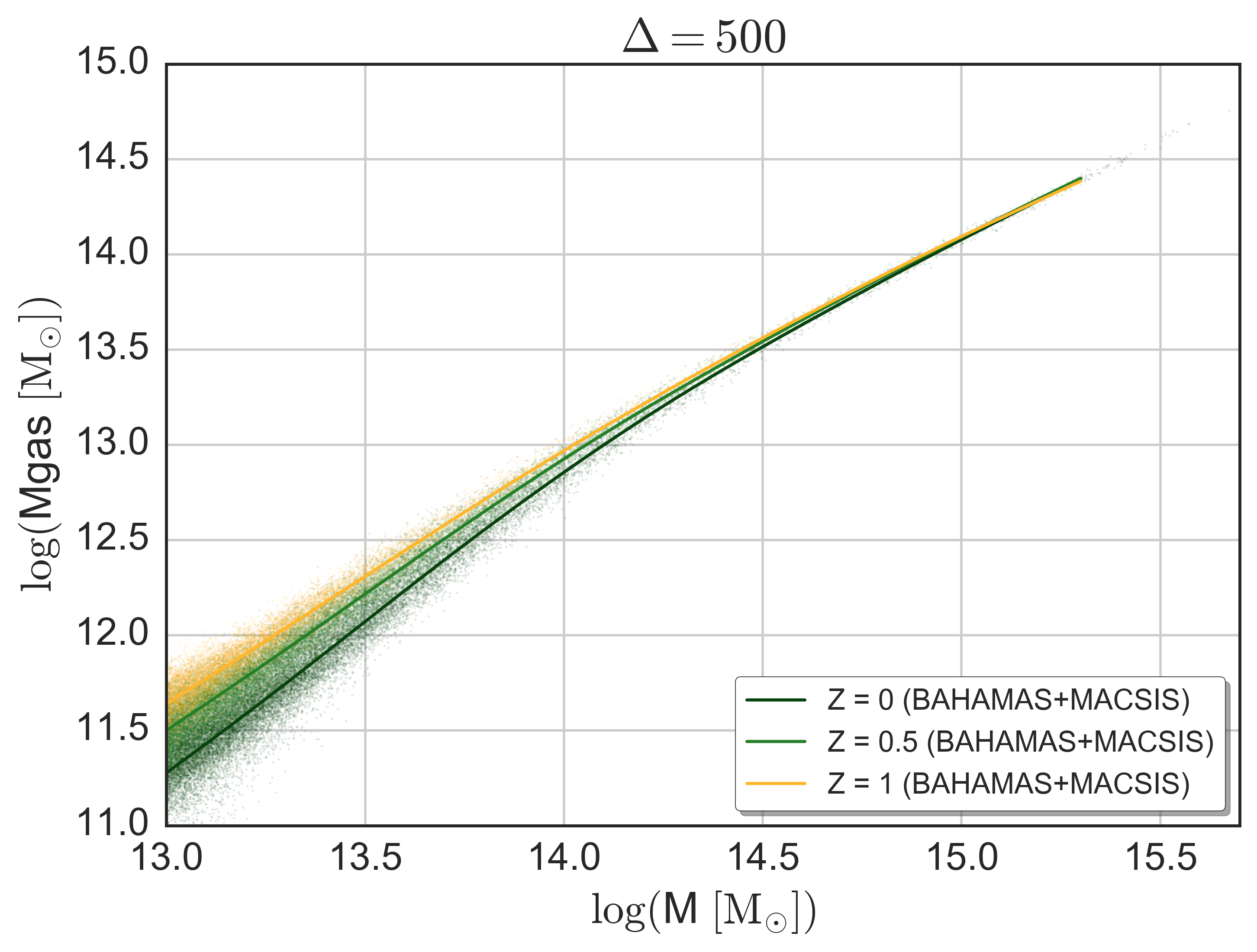}
  \end{subfigure} 
  \begin{subfigure}[b]{0.48\textwidth}
        \centering 
     	\includegraphics[width=\textwidth]{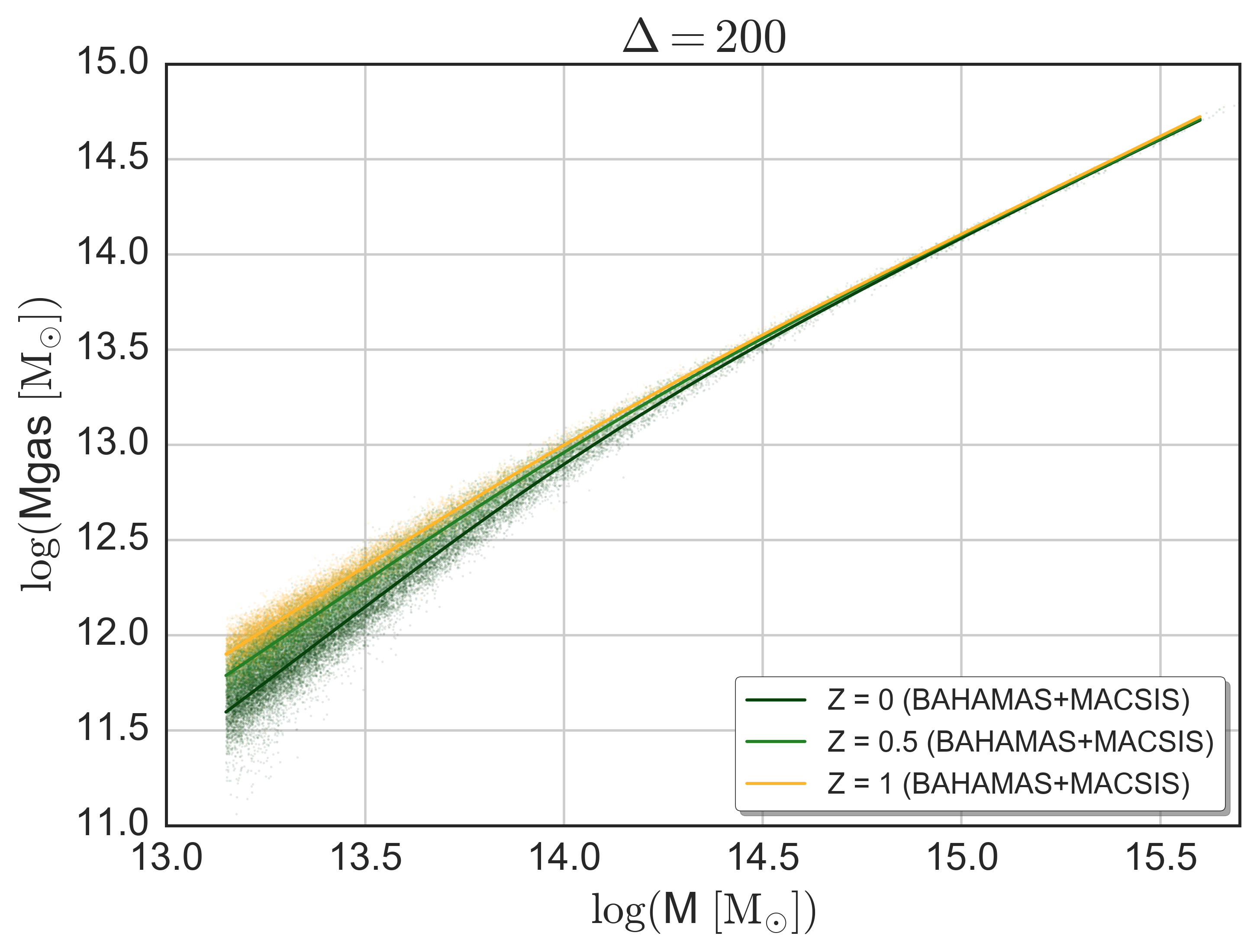}
  \end{subfigure} \\
   \begin{subfigure}[b]{0.48\textwidth}
       \centering 
    	\includegraphics[width=\textwidth]{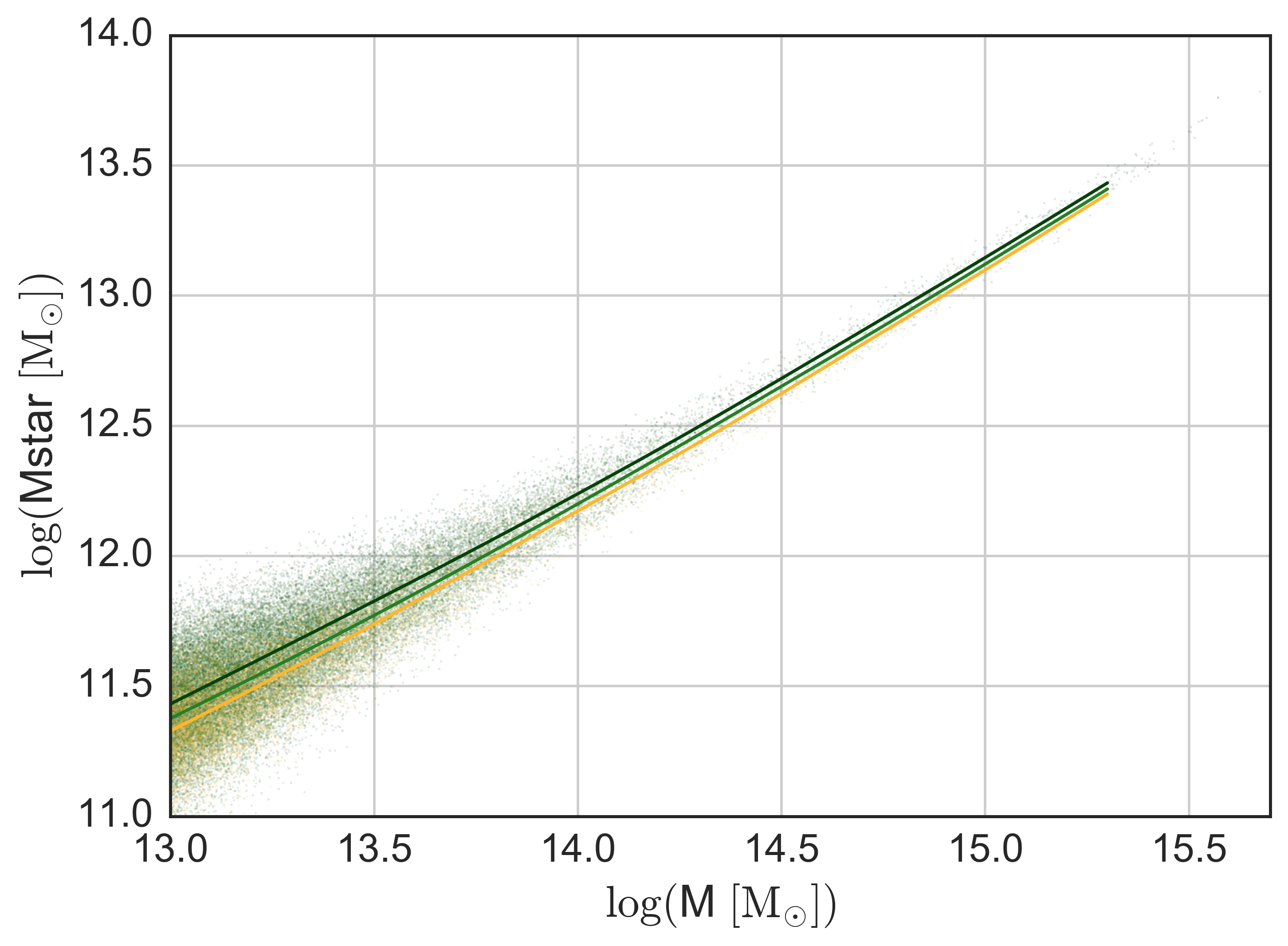}
   \end{subfigure} 
   \begin{subfigure}[b]{0.48\textwidth}
        \centering 
     	\includegraphics[width=\textwidth]{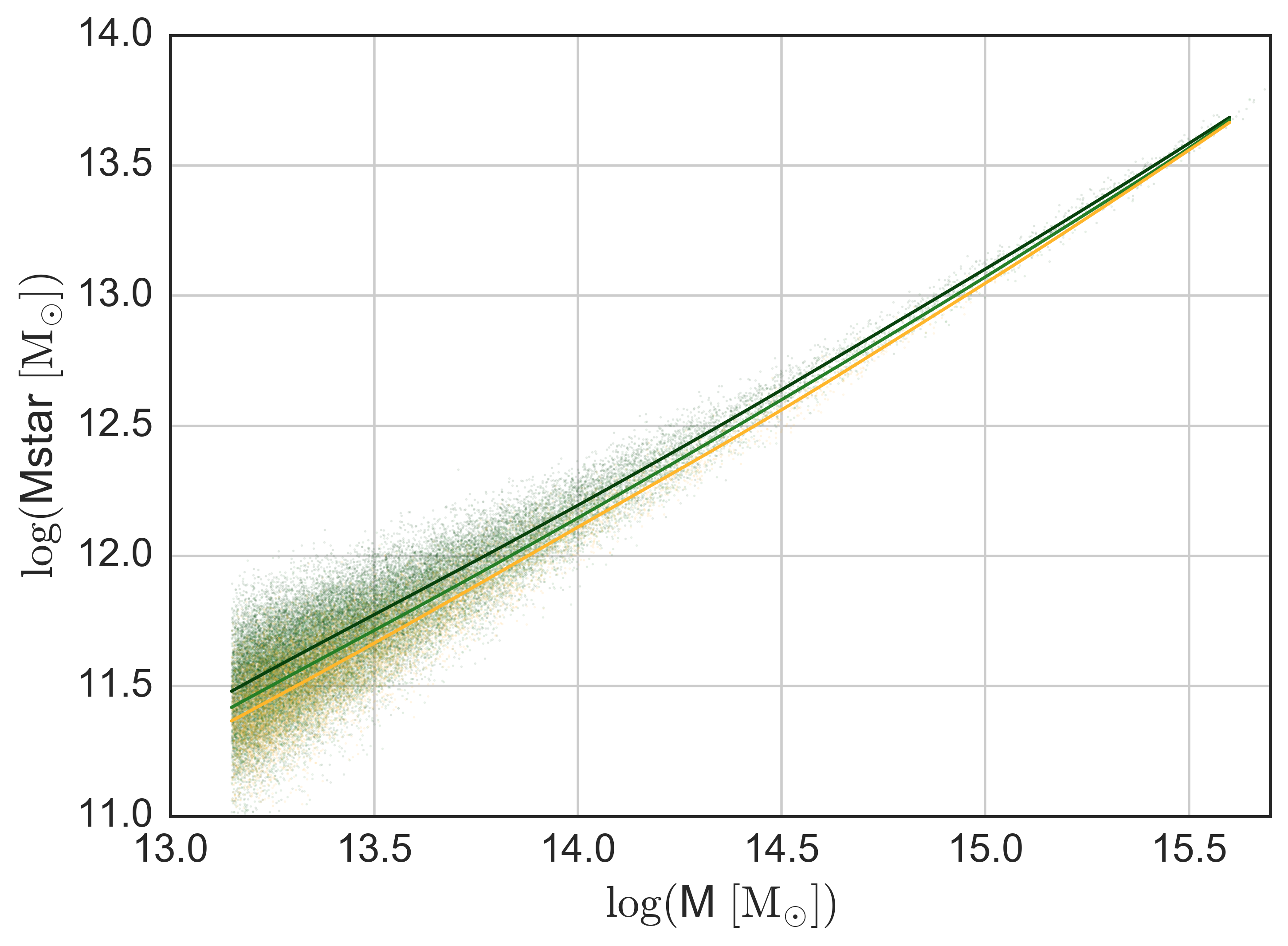}
   \end{subfigure} \\
   \caption{Halo baryon contents (points) measured within over-densities, $\Delta = 500$ (left) and $200$ (right), for $M_{\rm gas}$ (top) and $M_{\rm star}$ (bottom) as a function of total halo mass at three redshifts indicated in the legend. Lines show the LLR fits.  Parameters for the $\Delta = 500$ case are shown in Figures~\ref{fig:mass-evolution} and \ref{fig:fgas-star}. }
   \label{fig:scaling-relation}
\end{figure*}

In this section, we describe a localized linear regression model to characterize the conditional joint property likelihood, $p(\mstar,\ \mgas | \ \mhalo,z)$, of the simulated halo ensemble.  In practice, the power-law nature exhibited by most properties with respect to mass motivates the use of logarithmic variables that we introduce below.  

The method produces \emph{mass localized} estimates of the intercepts, slopes and covariance of this pair of properties as a function of halo mass at fixed redshift.  The assumption of a log-normal form for the conditional likelihood underlies this model, and we demonstrate the validity of this assumption in \S\ref{sec:residuals}.

 Following \citetalias{Evrard:2014}, our underlying population model considers a vector of properties, $\mathbf{S}$, associated with halos of total mass, $M_{\Delta}$, at redshift, $z$.  Using natural logarithms of the properties, $\mathbf{s} = \ln\mathbf{S}$, and mass, $\mu = \ln M_{\Delta}$, the log-mean scaling of property $a$ at a fixed redshift is locally linear
 \begin{equation} \label{eq:scalingmodel}
 \langle s_a \, | \, \mu, z \rangle \ = \ \pi_a(\mu,z) + \alpha_a(\mu,z) \mu , 
\end{equation}
with redshift- and scale-dependent parameters that we measure by differentially weighting halos in the simulation ensemble around a chosen mass scale. In this model the normalization of the property element, $S_a$, is $e^{\pi_a(\mu,z)}$. 

At a fixed redshift, we determine local fit parameters --- the slope $\alpha_a(\mu)$, intercept, $\pi_a(\mu)$, and intrinsic sample variance, $\sigma^2_a(\mu)$ --- for property $s_a$ by minimizing the weighted square error,
\begin{equation}\label{eq:squareError}
\epsilon^2_a(\mu) = \sum\limits_{i=1}^{n} \ w_i^2 \ (s_{a,i}- \alpha_a(\mu) \mu_i - \pi_a(\mu) )^2,
\end{equation}
where the sum $i$ is over halos, $\mu_i \equiv \ln (M_{{\rm halo},i}/M)$, and $w_i$ is the local weight centered on the mass scale, $M \equiv e^{\mu}$.  We sweep through values of $M$ covering the mass scale of poor groups to rich clusters, $M_{500} \in \{10^{13},10^{15}\} \msol$, in the joint \bahamas and \macsis halo samples.  

We use a Gaussian weight in log-mass, 
\begin{equation} \label{eq:wights}
 w_i \ = \ \frac{1}{\sqrt{2 \pi} \sigma_{\rm LLR} } \exp\left\{-\frac{\mu_i^2}{2\sigma_{\rm LLR}^2} \right\},
\end{equation}
with $\sigma_{\rm LLR} = 0.46$, equivalent to $0.2$ dex in halo mass. As the central halo filter scale, $\mu$, is varied, we record the local slope and intercept fit parameters.

With a local slope and intercept for each property, $j$, we can compute the local property covariance using the same weighting scheme. 
We use an unbiased weighted estimator of the property covariance matrix, $C$ \citep{GNU:2009}, 
\begin{equation} \label{eq:r-estimator}
C_{a,b} = A \sum\limits_{i=1}^{n}w_{i} ~ \delta s_{a,i} ~ \delta s_{b,i},
\end{equation}
where $\delta s_{a,i} \equiv s_{a,i}- \alpha_a \mu_i - \pi_a$ is the residual deviation from the local best-fit, $(a,b)$ are labels representing either stellar mass or hot gas mass, and the pre-factor is 
\begin{equation} \label{eq:A}
A = {\frac {\sum\limits_{i=1}^{n}w_{i}}{\left(\sum\limits _{i=1}^{n}w_{i}\right)^{2}-\sum\limits _{i=1}^{n}w_{i}^{2}}} .
\end{equation}

The covariance matrix for our pair of halo properties has one correlation coefficient, 
\begin{equation} \label{eq:A}
r_{\rm gas, star} = \frac{ C_{\rm gas,star}}{\sqrt{C_{\rm gas,gas} \ C_{\rm star,star}} }.
\end{equation}

We note that fitting a global power-law to MPRs that run with scale could induce covariance as an artifact of the poor, i.e. underfit, regression model.  
The locally estimated covariance is unbiased, easily computable, and asymptotically approaches the population true value in the limit of $\sigma_{\rm LLR} \rightarrow 0$ and $N_{\rm halo}\rightarrow\infty$.

\section{Results}\label{sec:results}

In this section, we begin by presenting the LLR scaling behavior of log-mean stellar mass and hot gas mass as a function of halo mass and redshift.  We then examine the form of the conditional likelihood PDF, finding excellent agreement with a log-normal form, the assumption behind the weighted Pearson covariance, equation~(\ref{eq:r-estimator}). Finally, we investigate the redshift and mass dependence of the star-gas covariance. 

Unless otherwise stated, error bars and shaded regions in the figures below are one standard deviation based on bootstrap estimates of 1000 re-sampled halo datasets.

\subsection{LLR fits to scaling relations}

Figure~\ref{fig:scaling-relation} shows how the hot gas mass (top) and stellar mass (bottom) of the \bahamas and \macsis halo population scale with total mass at three redshifts and for two critical overdensity scales, 
$\Delta = 500$ and 200.  LLR fit lines are also shown. Overall, the conditional statistics display similar forms at different overdensities and redshifts, but the fit parameter values depend on scale, redshift and halo mass.  

\begin{figure} 
  \begin{subfigure}[b]{0.49\textwidth}
       \centering 
    	\includegraphics[width=\textwidth]{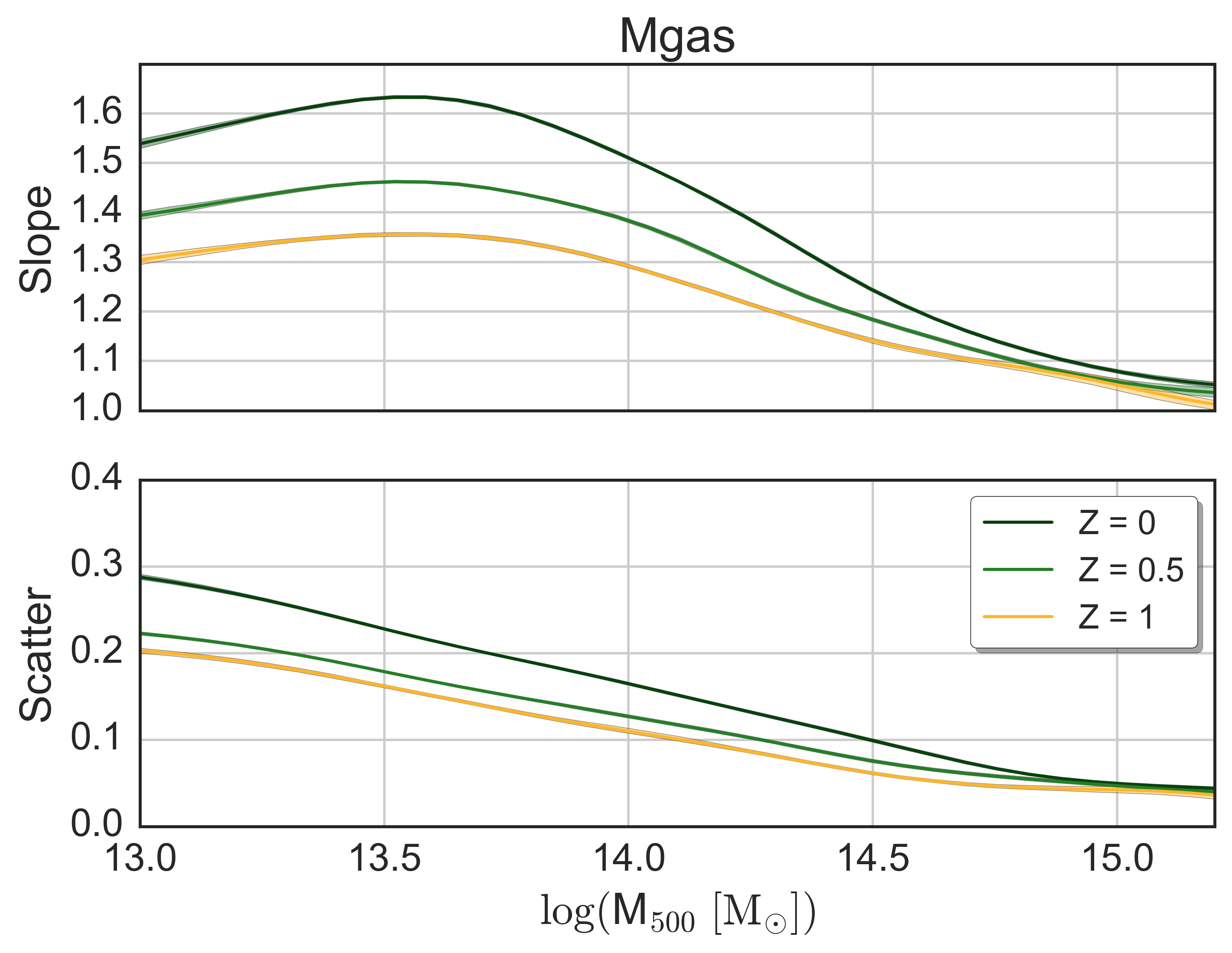}
  \end{subfigure} \\
   \begin{subfigure}[b]{0.49\textwidth}
       \centering 
    	\includegraphics[width=\textwidth]{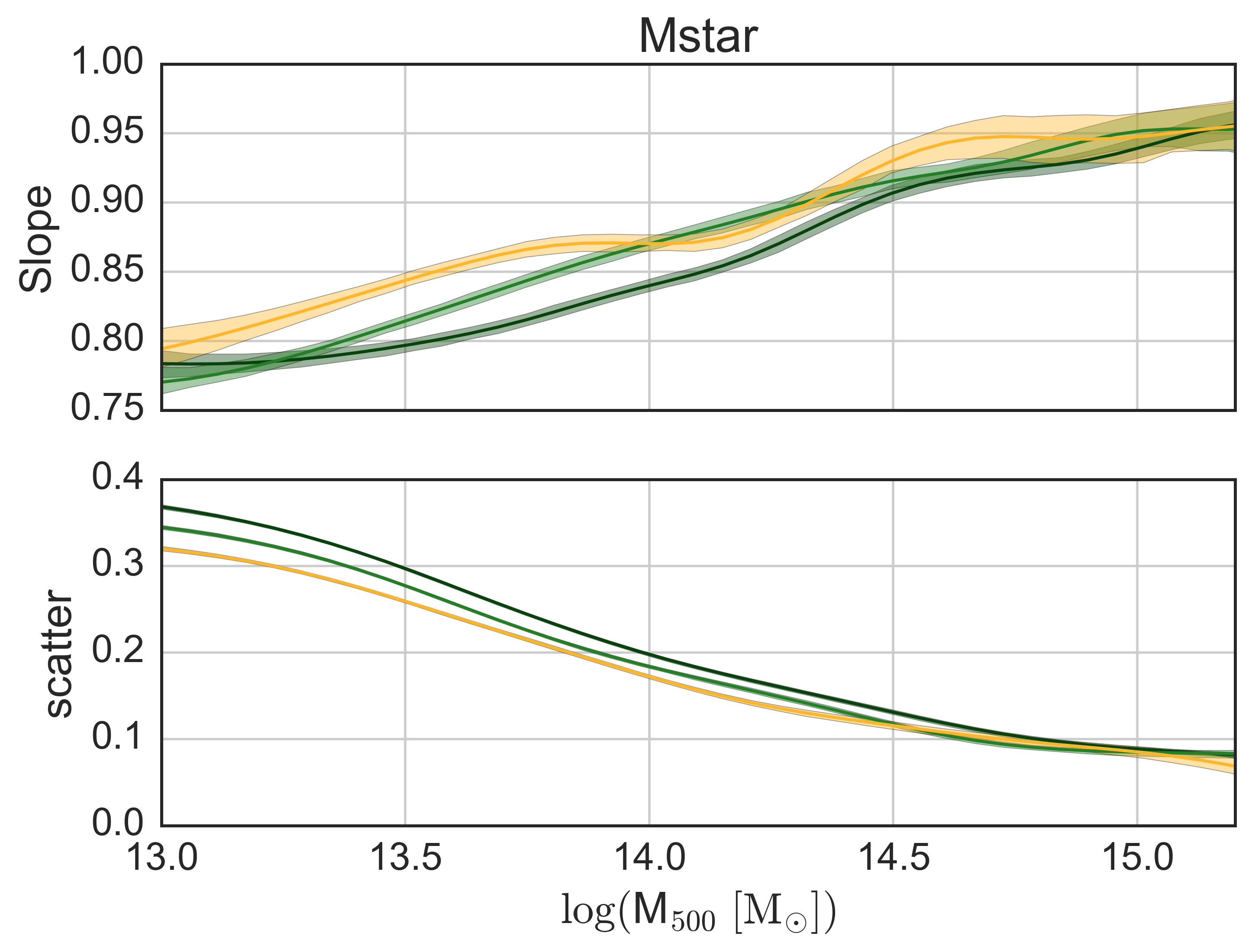}
   \end{subfigure} \\
   \caption{Dependence of the slope and scatter of hot gas mass (top) and stellar mass (bottom) MPRs on total halo mass for $\Delta =500$. Lines show the LLR estimates and shaded regions give $1\sigma$ confidence bootstrap errors in the parameters.  The scatter is the root-mean square of the natural log. }
   \label{fig:mass-evolution}
\end{figure}

 Figure \ref{fig:mass-evolution} shows the mass and redshift dependence of the gas/star LLR slope and rms scatter at $\Delta = 500$. There is strong scale dependence in the slopes of the MPR scalings in both $\mgas$ and $\mstar$, with milder redshift dependence.  For $\mgas$ both the slope and scatter at fixed halo mass increase at lower redshifts, and the running behavior of the slope is non-monotonic with halo mass, exhibiting a peak value near a group-scale mass, $M_{500} \sim 3\times 10^{13} \msol$.  For $\mstar$ the redshift sensitivity of the MPR parameters at fixed halo mass is more modest, and the slope at tends to slightly decrease toward lower redshifts.  The running of the $\mstar$ slope is approximately linear in the log of halo mass.

In the \bahamas simulation study of \citet{LeBrun:2016}, a broken (piece-wise constant) power-law is used to fit the scaling of hot gas mass with halo mass. The broken power-law approach introduces a particular mass scale --- the transition, or break, mass --- that is not anticipated by the relatively smooth astrophysical processes operating within halos.  The LLR approach enables the detection of continuously varying, scale-dependent features without introducing an arbitrary halo mass scale.  Indeed, the smooth behaviors of the local slopes in Figure \ref{fig:mass-evolution} do not support a broken power-law approximation for either hot gas mass or stellar mass. 

\begin{figure} 
   \centering 
   \includegraphics[width=0.49\textwidth]{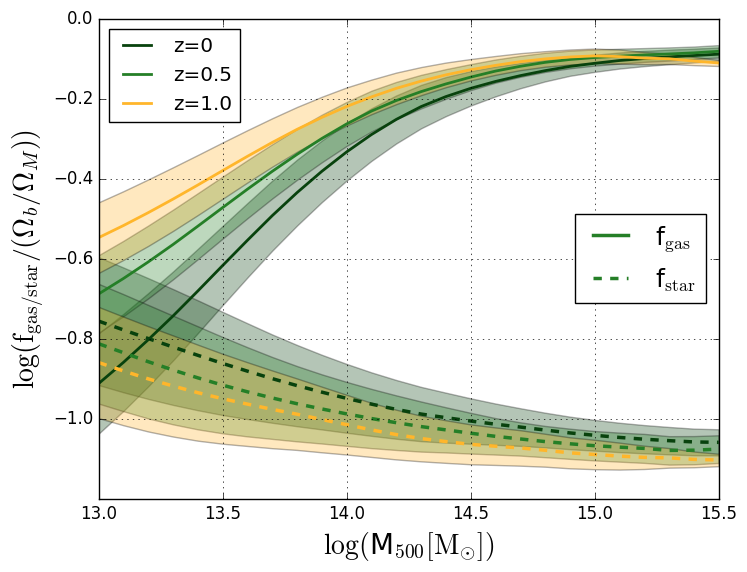}
   \caption{LLR normalizations of hot gas mass (solid) and stellar mass (dashed), expressed as mass fractions, $f_a = e^{\pi_a(\mu,z)}/M$, where $\pi_a(\mu,z)$ is the scale- and redshift-dependent log-mean, equation~(\ref{eq:scalingmodel}), normalized by the cosmic mean baryon fraction of the \bahamas universe.  Shaded regions show the intrinsic scatter within the population rather than uncertainty in the mean behavior. }
   \label{fig:fgas-star}
\end{figure}

For cluster-scale systems above $\sim 5 \times 10^{13} \msol$, the slopes in both gas mass and stellar mass run nearly linearly with log-mass, approaching the naive self-similar expectation of one in the highest mass systems from above and below, respectively.  This is in agreement with \citet{Barnes:2017} who find a slope $\sim 1$ when only the most massive systems are considered, but find a steeper slope using the superset of \bahamas and \macsis halos more massive than $10^{14}~\msun$.  

 As hierarchical clustering progresses and halos grow larger and develop deeper potential wells, feedback driven by the central galaxy becomes more confined to the core region, allowing gravity to become dominant and self-similar scalings to recover. The simulations show this type of progression, with slopes at $z=0$ in $\mgas$ and $\mstar$ lying within $1.00 \pm 0.05$ at masses, $M_{500} > 10^{15}~\msun$. 
Furthermore, for the highest-mass systems, the MPR parameters do not vary significantly with redshift, but there are statistically significant changes in the  slope and normalization for group-scale systems.  

The above trends persist at both overdensity scales presented in this work.  We confirm, but do not present here, similar behavior at $\Delta = 2500$. The LLR fit parameters for $\Delta = 500$ and $200$ are provided in Appendix~\ref{app:raw-data}.  

 Figure~\ref{fig:fgas-star} shows the scale and redshift behavior of the $\Delta = 500$ LLR normalizations for stellar and hot gas masses.  The normalizations are presented as halo mass fractions normalized by mean cosmic baryonic fraction.  Recall that we have aligned the \macsis cosmic baryon fraction to that of the \bahamas simulation.   

Above a halo mass of $\sim 3 \times 10^{14} \msol$, the total gas mass and stellar mass fractions become nearly constant; however, there is strong mass and redshift evolution for lower mass systems.  The nearly fixed high mass behavior provides strong evidence that baryon venting is negligible, while considerable venting occurs at the mass scale of groups.  The weak redshift dependence at high mass is in good agreement with trends observed from a joint analysis of South Pole Telescope (SPT) and Dark Energy Survey (DES) data in a sample of 93 massive SPT clusters \citep{Chiu:2017}.

 The interplay between cooling and feedback controls the relative mean proportions of the integrated gaseous and stellar masses in a way that introduces considerable variance at the group mass scale, but the variance decreases for richer clusters with deeper potential wells.  Associated with this, the covariance of gas and stars determines the scatter in overall baryon content. We find evidence for a  ``closing box'' scenario at the high-mass end, with increasing anti-correlation of stellar mass and gas mass at later times.  We present this result in Section~\ref{sec:covar}. 

\subsection{Log-normality of conditional statistics} \label{sec:residuals}

\begin{figure}
  \centering
  \setbox1=\hbox{\includegraphics[height=2.3cm]{example-image-b}}
  \includegraphics[width=0.49\textwidth]{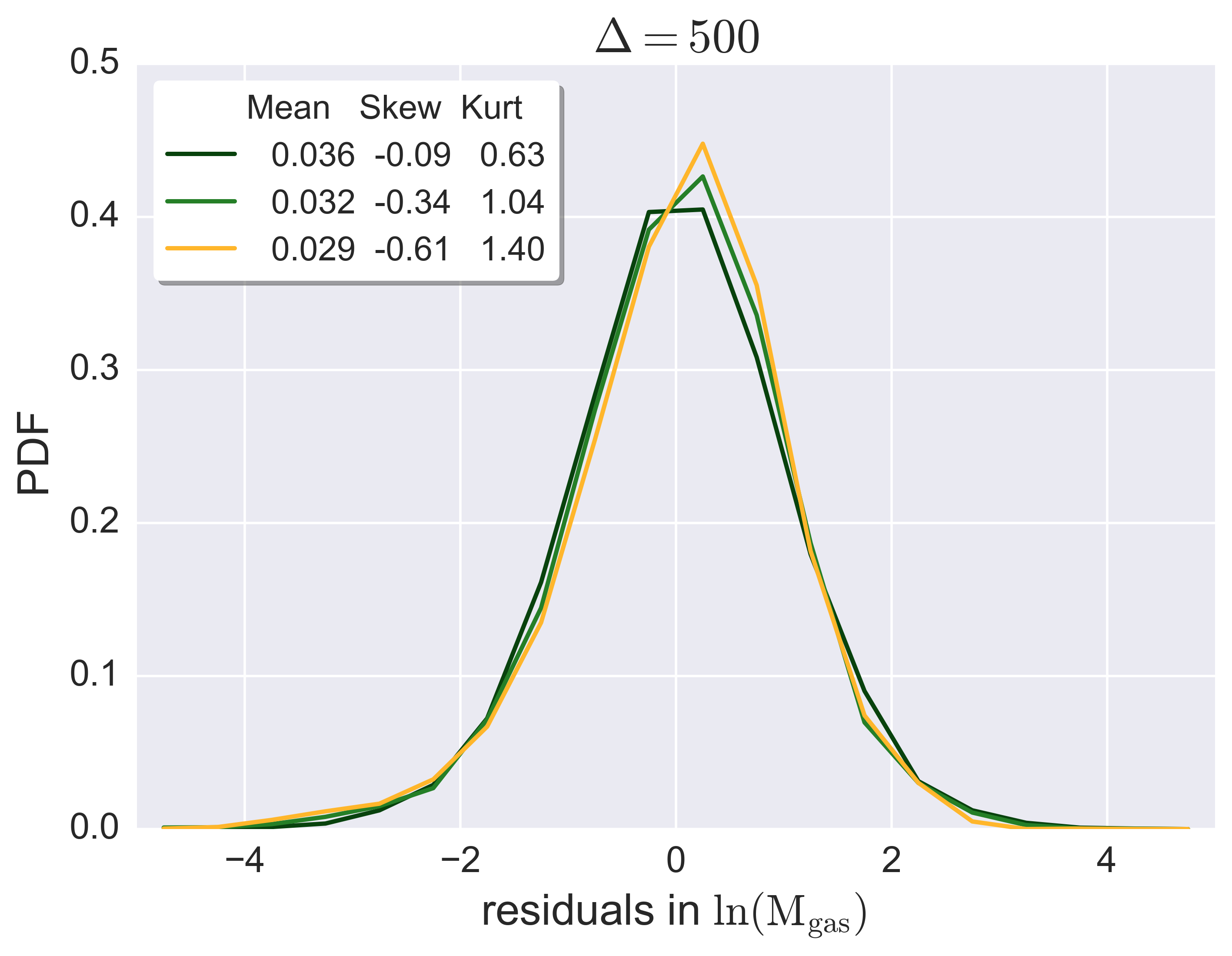}\llap{\makebox[\wd1][l]{\raisebox{3.9cm}{\includegraphics[height=2.3cm]{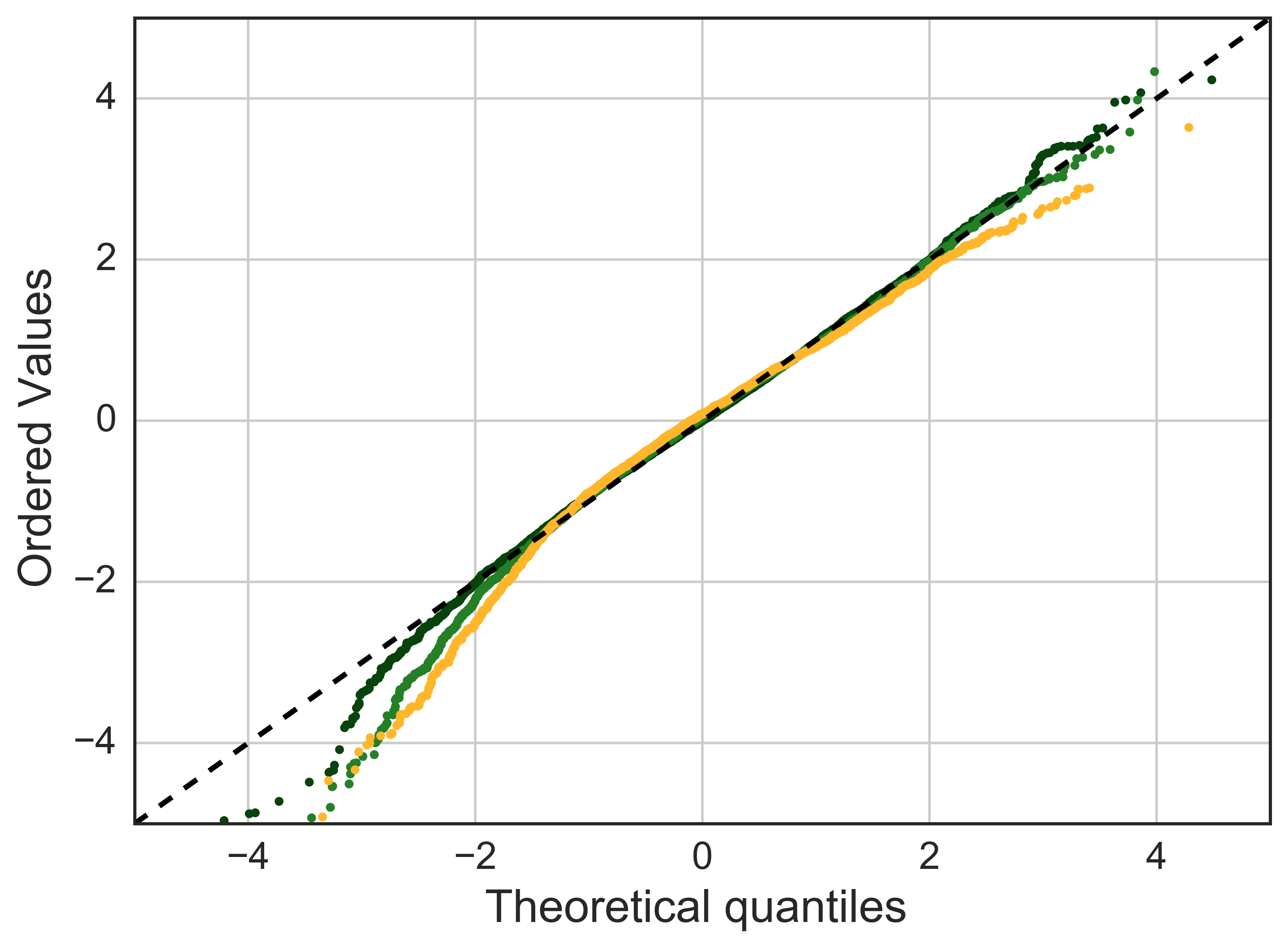}}}} \\
  \includegraphics[width=0.49\textwidth]{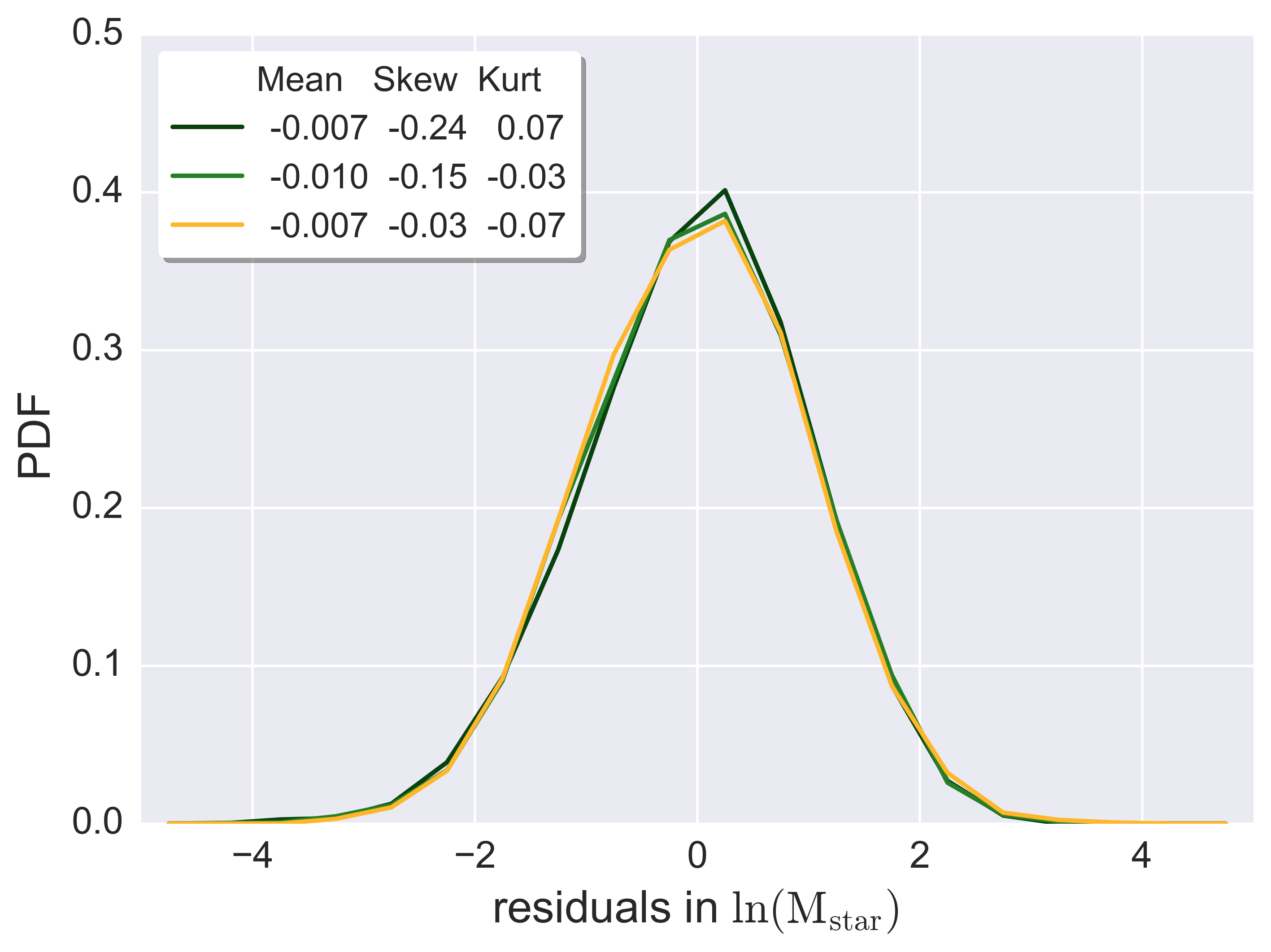}\llap{\makebox[\wd1][l]{\raisebox{3.9cm}{\includegraphics[height=2.3cm]{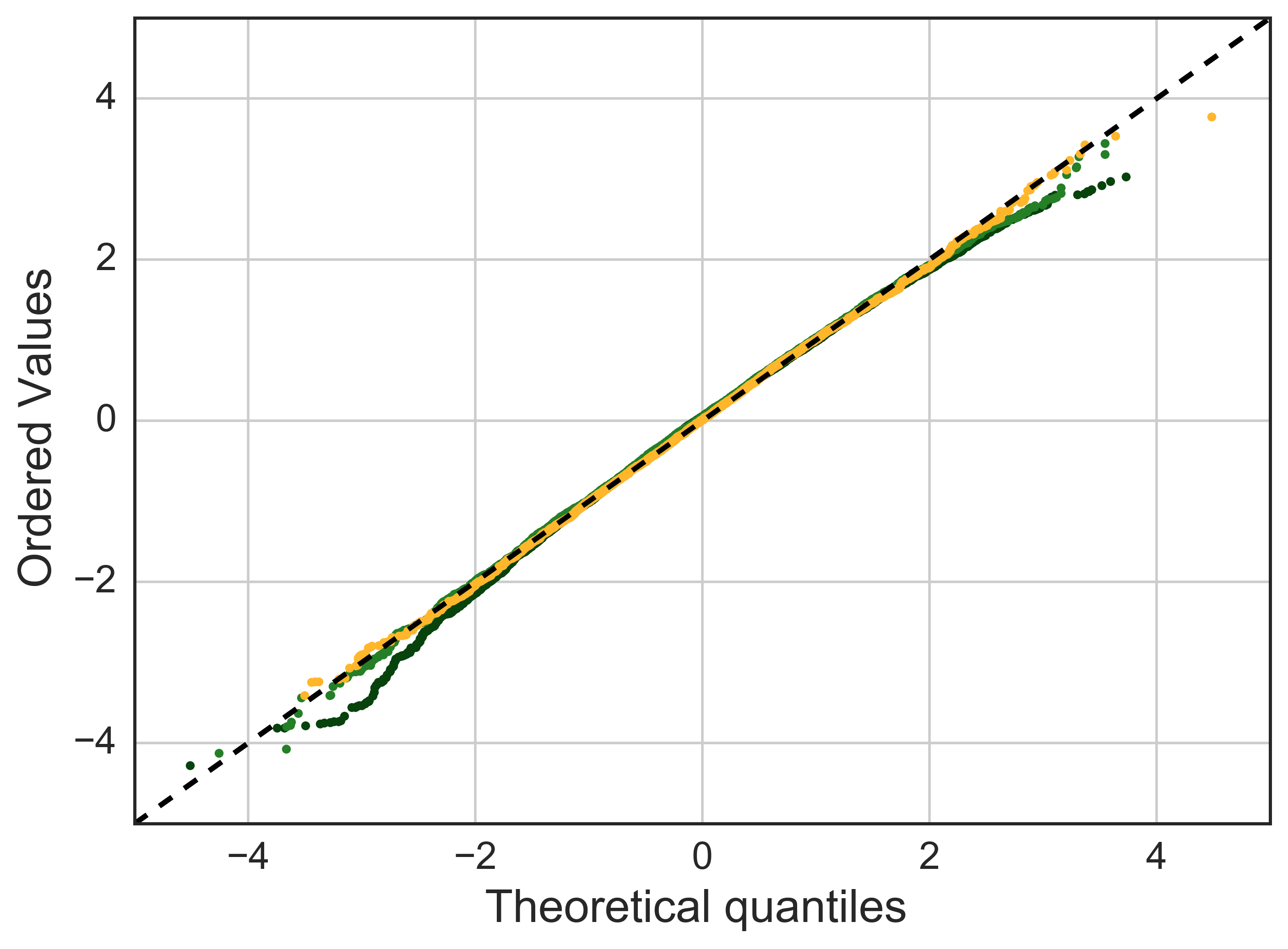}}}}
  \caption{Conditional likelihood distribution derived from scaling relation residuals, equation~(\ref{eq:residual}) in hot gas mass (top) and stellar mass (bottom). Colors indicate redshift as in Figure~\ref{fig:scaling-relation}. The mean bias is typically less than $1\%$, skewness is less than $1$, and kurtosis is less than $5$ which are strong indicators of log-normality.  Rank (Q-Q) comparison, shown in the inset of each panel, indicate only mild deviations in log-normality in the wings of each distribution. } \label{fig:residuals}
\end{figure}

The log-normal shape of conditional statistics, an implicit assumption in previous analyses, is a core ingredient of the \citetalias{Evrard:2014} population model.  In the context of modeling star formation, a log-normal shape for final stellar masses is expected when random multiplicative factors govern the evolution of the system \citep[e.g.][]{Larson:1973, Adams:1996}.  Observational studies of galaxy clusters broadly support this form, although with currently modest sample sizes \citep[e.g.][]{Pratt:2009,Mantz:2010,Czakon:2015,Mantz:2016}.  

Non-Gaussian terms in MPR statistics can introduce bias in cosmological analysis based on cluster counts \citep{Erickson:2011,Weinberg:2013}.  Such terms cannot be characterized through measurement of the scatter alone.  We use the large \bahamas halo samples to study the PDF shape in detail, and assess the degree to which conditional property statistics of the simulated halo sample follow a log-normal frequency distribution. 

Previous simulation studies have addressed this issue with generally smaller samples.  Using an ensemble of N-body and non-radiative hydrodynamics simulations, \citet{Evrard:2008} show that the PDF of dark matter velocity dispersion at fixed halo mass is very close to log-normal, with some samples showing a modest skew caused by a minority population of post-merger, transient systems. The construction of the \bahamas and \macsis halo samples effectively filters out the small fraction of such secondary objects.   
\citet{Stanek:2010} demonstrate log-normal PDFs for multiple properties within a sample of $\sim 4000$ halos drawn from the Millennium Gas Simulations, as do other hydrodynamic simulations with smaller samples  
\citep{Fabjan:2011,Biffi:2014,LeBrun:2016,Truong:2016}.

Given the LLR fit for property $s_a$ (with $a$ a label indicating either $\ln \mstar$ or $\ln \mgas$), we calculate the normalized deviation of halo $i$ from the mean relation, 
\begin{equation}\label{eq:residual}
\tilde{\delta}_{a,i} \equiv \delta s_{a,i} /\sigma_a(\mu_i) = \frac{s_{a,i}- \alpha_a(\mu_i) \mu_i - \pi_a}{\sigma_a(\mu_i)}, 
\end{equation}
where $\alpha_a(\mu_i)$ and $\sigma_a(\mu_i)$ are the local slope and scatter of the MPR evaluated at the total mass of the $i^{\rm th}$ halo (see, Figure~\ref{fig:mass-evolution}).

Figure~\ref{fig:residuals} presents the PDF of the normalized residuals of gas mass (top panels) and stellar mass (bottom panels) for $\Delta=500$ at $z=0$, $0.5$ and 1. These results are consistent for all overdensities.  The inset of each panel provides a Q-Q plot\footnote{The quantile-quantile (Q-Q) plot is a visualization technique for determining if a population sample comes from an assumed distribution. Axes compare rank quantiles of the model to quantiles of the sample.} 
to illustrate deviations from the normal form.  The residuals in the log of stellar mass are extremely Gaussian, while the gas mass displays slight negative skewness and non-zero kurtosis.  We note that only a small fraction halos, $< 1\%$, are outliers with low gas mass.  Understanding the physical causes of this minor deviation from normality lies beyond the scope of this work.  
The Gaussian form persists for both $\mgas$ and $\mstar$ and over all over-density scales considered in this work.

These results provide strong evidence that the log-normal form is adequate to model the {\sl intrinsic} quantities of halos. 
In Section~\ref{sec:E14-test} we demonstrate that employing a local form of the \citetalias{Evrard:2014} model achieves sub-percent accuracy in estimating the population mean mass selected on baryon mass.

Within the scope of cluster cosmology, non-Gaussian MPR shapes were formulated by \citet{Shaw:2010} in terms of an Edgeworth series expansion, 
\begin{equation}
    P(M_{\rm proxy} | M_{\rm true}) \approx G(x) - \frac{\gamma}{6}\frac{{\rm d}^3G}{{\rm d}x^3} + \frac{\kappa}{24}\frac{{\rm d}^4G}{{\rm d}x^4} + \frac{\gamma^2}{72}\frac{{\rm d}^6G}{{\rm d}x^6},
\end{equation}
\noindent where the skewness, $\gamma$, is defined as, 
\begin{equation}
\gamma = \frac{\langle (M_{proxy} - M_{true})^3 \rangle}{\sigma^2},
\end{equation}
\noindent and the kurtosis, $\kappa$, is defined as,
\begin{equation}
\kappa = \frac{\langle (M_{proxy} - M_{true})^4 \rangle}{\sigma^4} - 3,
\end{equation}
\noindent and $G(x)$ is a Gaussian distribution.  We note that achieving sub-percent level systematic uncertainty in  cluster number counts under a log-normal approximation with a mass proxy having $20\%$ scatter requires roughly $\gamma < 7$ and $\kappa < 90$ \citep[see, equation (156) of][]{Weinberg:2013}. The skewness and kurtosis values for our halo samples are at least an order of magnitude smaller than what is needed to achieve sub-percent uncertainty in number count statistics, but more work is needed to confirm this result for realistic cluster samples.

In principle, if the form of an observable conditional statistics at fixed halo mass is known, it can be easily incorporated into a cosmological analysis without introducing additional source of systematic error due to the uncertainty in the form of distribution.  When modeling observational data, the form of the conditional statistics of measured quantities may differ from a log-normal form, for example due to projection effects \citep[{\sl e.g.}, ][]{Cohn:2007, Erickson:2011}. 
Analysis of such data using a log-normal assumption in the likelihood leads to systematic biases in halo mass that in turn can bias cosmological parameter constraints.
These additional uncertainties are strongly dependent on survey characteristics and data reduction pipeline and so must be modeled explicitly \citep[{\sl e.g.},][]{Juin:2007,Farahi:2016,Pacaud:2016,deHaan:2016}.  

\subsection{Stellar--hot gas covariance}\label{sec:covar}

A complete multi-wavelength MPR likelihood model will include property covariance.  
For cosmology, knowledge of property covariance improves dark energy constraints when performing analysis of joint, multi-wavelength cluster samples \citet{Cunha:2009}. For astrophysical studies, \citet{Nord:2008} demonstrate how covariance between temperature and luminosity can confuse studies of luminosity-temperature redshift evolution.  Covariance of observed hot gas properties has recently been measured in X-ray selected samples \citep{Mantz:2010,Mantz:2016,Andreon:2017}.  

In simulations, a covariance matrix of dark matter and hot gas properties was first presented by \citet{Stanek:2010} for halo samples in the Millennium Gas simulation. Based on a small sample of high mass halos and their progenitors run with RAMSES hydrodynamics including AGN feedback, \citet{Wu:2015} published the first non-zero correlation of hot gas and stellar mass fractions.  We perform a similar measurement here on a much larger sample of halos evolved with an independent numerical method.  

\begin{figure}
   \includegraphics[width=0.48\textwidth]{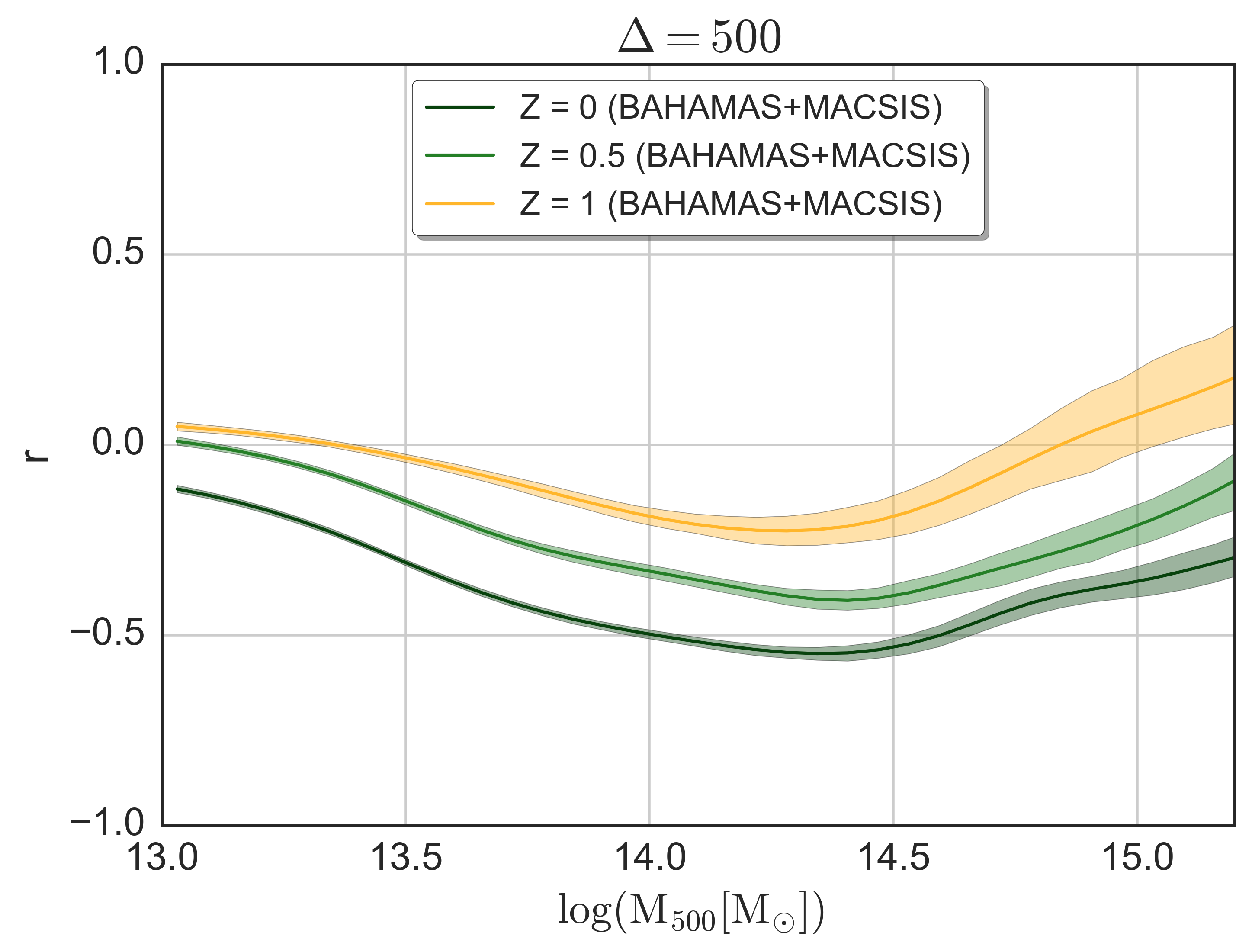}
   \caption{The LLR correlation coefficient between stellar mass and gas mass at fixed halo mass, equation \ref{eq:r-estimator} at the redshifts indicated. Anti-correlation is favored at low redshifts and masses above $10^{14} \msol$.}
   \label{fig:corr}
\end{figure}

The correlation coefficient of gas and stellar mass at fixed total mass, equation \ref{eq:r-estimator}, is plotted as a function of halo mass in Figure~\ref{fig:corr}.  The color scheme is consistent with that used in Figure~\ref{fig:scaling-relation}.  
The correlation coefficient begins near zero at $10^{13} \msol$ and becomes increasingly negative at higher halo mass. The values plateau around $3 \times 10^{14} \msun$ and decline in amplitude for the highest mass halos. While we show the results at $\Delta=500$, the pattern at $\Delta=500$ is similar.  

The lack of correlation for group size halos can be explained through an ``open box'' scenario in which the total baryonic content of a halo is not conserved.   Feedback effects at low masses are efficient at venting material out of the relatively shallow potential well.  As shown by \citet{McCarthy:2011}, the gas ejection takes place at high-redshifts, $2 \la z \la 4$, in the progenitors of present-day groups.  The ejection is sufficiently energetic that the gas is not re-accreted later on.  For higher mass halos, however, the gas is re-accreted.  The anti-correlation above $10^{14} \msol$ is indicative of a more ``closed box'' nature in which the overall baryon fraction of halos more closely resembles the global value, $\Omega_b / \Omega_m$.  The redshift behavior in Figure~\ref{fig:corr} indicates that the box is closing more tightly over time, with the extremal value of $r$ decreasing from $-0.25$ at $z=1$ to $-0.5$ at $z=0$.  

\citet{Wu:2015} find a correlation coefficient of $-0.68$ at $\Delta = 500$, stronger than what is found here. 
The different behaviors appears are likely due to the smaller variance in stellar mass in the \bahamas and \macsis samples for the most massive systems, $\gtrsim 10^{15} \msun$. 
 We return to this issue in more detail in Section~\ref{sec:discussion}.

\section{Validating the analytic population model } \label{sec:E14-test}

Cluster population statistics are linked to the constituents of the universe through the growth of cosmic structure, and many ongoing and future cluster surveys are focused on using cluster population statistics to constrain models of dark energy and cosmic acceleration \citep[{\sl e.g.},][]{Mantz:2015,deHaan:2016,Mantz:2016,DES-DE+:2016,Pierre:2016}.  
The multi-property space density and conditional statistics of the population of massive halos are essential ingredients of such efforts.  The evidence presented above indicates that the \bahamas and \macsis halo populations obey the log-normal statistics assumed by the \citetalias{Evrard:2014} analytic model.  In this section we explicitly test the accuracy of that model by examining the expected log-mass of halos, $\langle \ln M | s_a \rangle$, selected by an intrinsic property, $s_a$. 

The mean, comoving number density of halos expected within some specific property bin, $i$, at redshift, $z$, is given by the convolution, 
\begin{equation}
\left\langle \frac{d n_i(z)}{d V} \right\rangle = \int_{s_i}^{s_{i+1}} \, ds  \, \int_{-\infty}^{\infty} d\mu \ \MFmu \ p(s | \mu,z), 
\label{eq:Nbar}
\end{equation} 
with $p(s|\mu,z)$ the conditional likelihood of the property used to select the halo sample, and $\MFmu$ is the mass function. 

The smoothness of the mass function allows a logarithmic polynomial expansion, 
\begin{equation}
\MFmu =  \exp \left[  \beta_0(z) - \sum_{j=1}^3 \ \frac{\beta_j(z)}{j !} \ \mu^j  \right] , 
\label{eq:MFmodel}
\end{equation} 
consisting of an amplitude, $e^{\beta_0(z)}$ and linear through cubic coefficients, $\beta_j(z)$, that control the shape.  These coefficients vary smoothly with redshift.  

\begin{figure}
\centering
\includegraphics[width=0.95\linewidth]{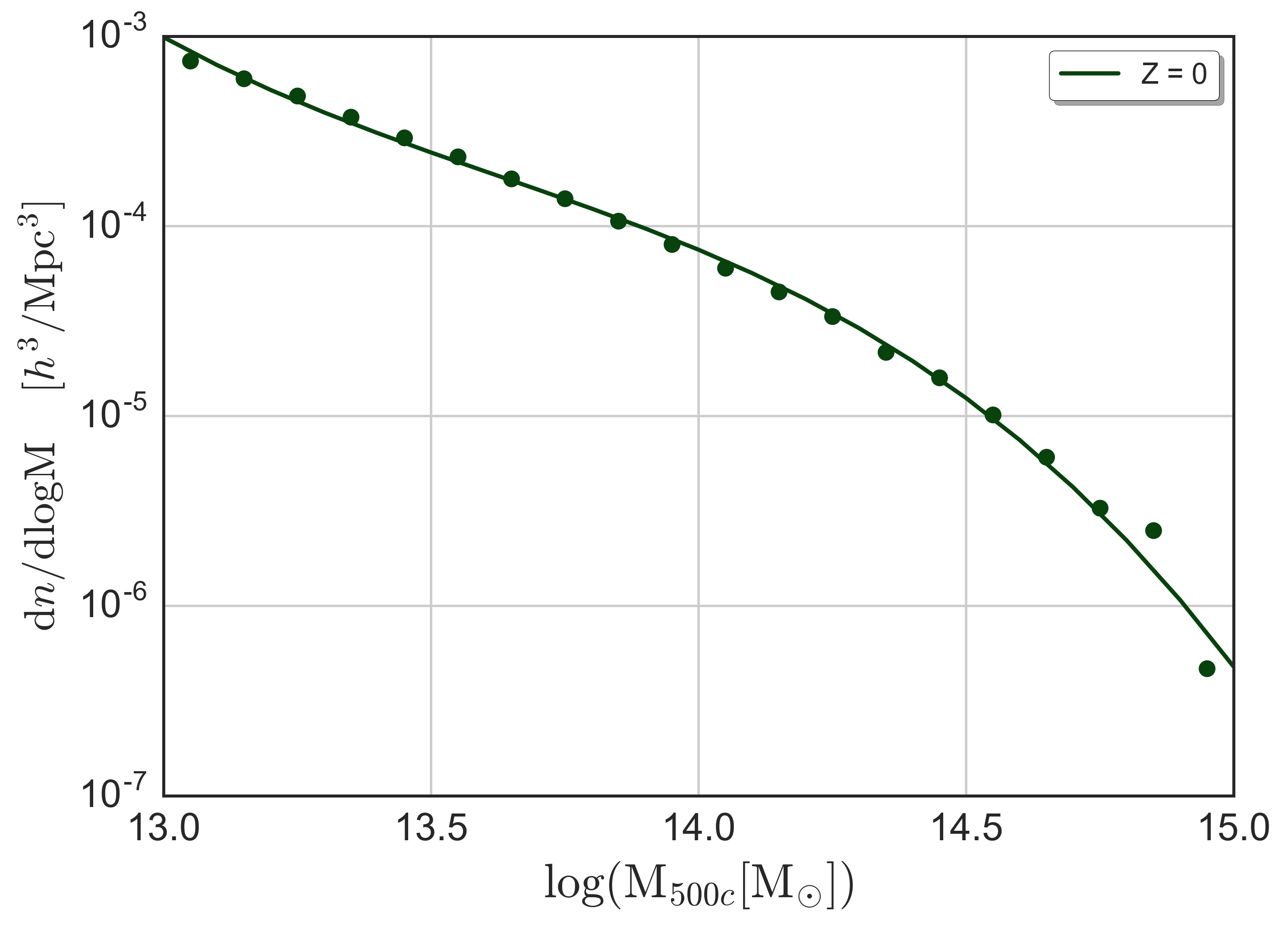} 
\caption{The halo mass function derived from the \bahamas simulation. The line is a third-order polynomial fit to the data points, equation~(\ref{eq:MFmodel}), for redshift $z=0$. }
\label{fig:MF}
\end{figure}

We analyze the $z=0$ sample and fit the number counts of halos to the above third-order polynomial. 
Figure~\ref{fig:MF} shows the differential number counts as a function of halo mass for redshift $z=0$ slices as points, and the corresponding mass function fits as lines.  Note that the $\beta_1$ and $\beta_2$ terms in \citetalias{Evrard:2014} are the {\sl local} first and second derivatives of \ref{eq:MFmodel} evaluated at a pivot mass, while the $\beta_1$ and $\beta_2$ in this work are derived from fitting the halo mass function over the mass range shown in Figure~\ref{fig:MF}. We find values of $\beta_0=8.42$, $\beta_1=2.93$, $\beta_2=0.86$, and $\beta_3=0.42$.

The convolution, equation~(\ref{eq:Nbar}), brings the halo mass function coefficients into the expression for the log-mean total halo mass selected by a given observable, $s_a$, 
\begin{equation}
\langle \mu \, | \, s_a, z \rangle =  x_s \, 
\left[ \left( \frac{s_a - \pi_a}{\alpha_a} \right) - \beta_1 \varmassfullone \ \right] ,
\label{eq:muGivenSa}
\end{equation} 
\noindent
where $\varmassfullone= \sigma_a^2/\alpha_a^2$ is the first-order estimate of the mass variance selected by property $s_a$, and
\begin{equation}
x_s \equiv (1 + \beta_2 \, \varmassfullone)^{-1} \simeq (1 - \beta_2 \, \varmassfullone), 
\label{eq:compression}
\end{equation} 
is a compression factor less than unity that is sensitive to the curvature of the mass function.  The $\beta_1$ term represents Eddington bias from convolution of a pure power-law mass function.  Generally, the slope of the mass function lies in the range $\beta_1 \in [2, 4]$, the curvature term $\beta_2 \simeq 1$, and the variance ranges from $(0.05)^2$ to $(0.3)^2$ (see Fig.~\ref{fig:mass-evolution}).

\begin{figure}
\centering
\includegraphics[width=0.95\linewidth]{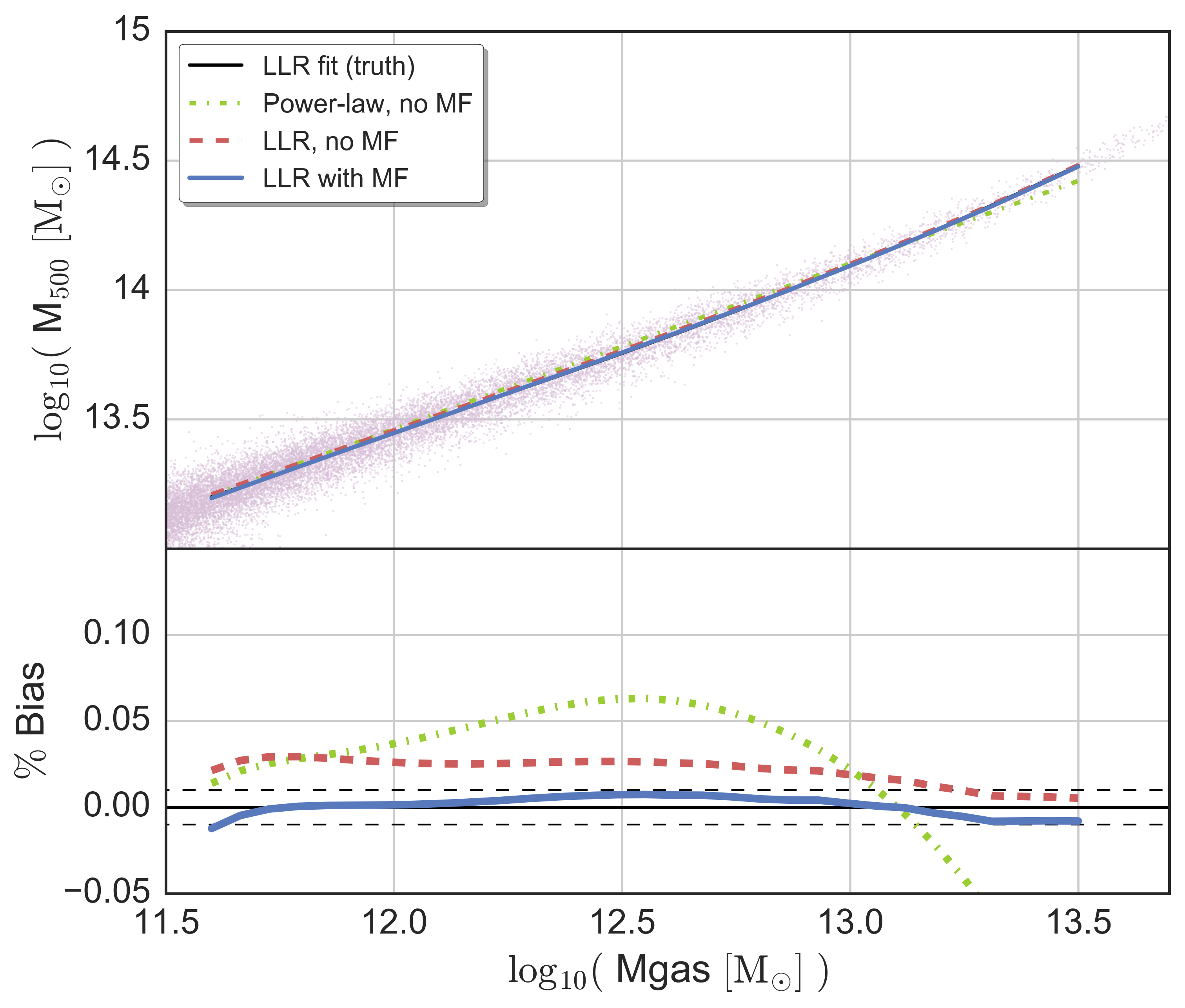} \\
\includegraphics[width=0.95\linewidth]{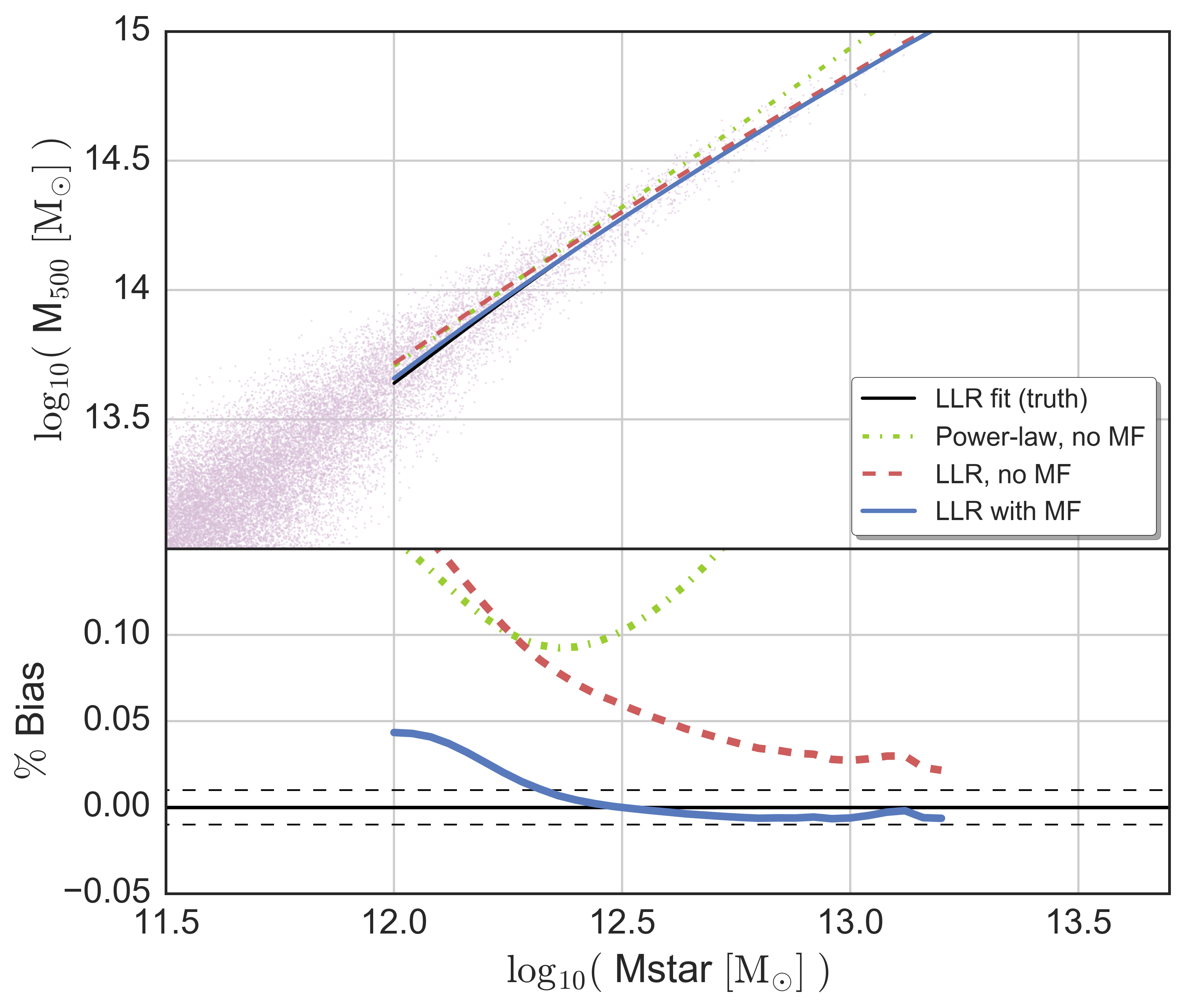}
\caption{Tests of the \citetalias{Evrard:2014} model for halos selected by hot gas mass (upper) and stellar mass (lower).  In each panel the upper sub-panels show the total halo mass of individual halos as a function of the selection mass, with black curves showing the LLR estimates of the underlying true $\langle \ln M_{500} | s_a \rangle$ relation, where $s_a = \ln \mgas$ or $\ln \mstar$. The red dashed (green solid) lines are predictions from inverting the global (local) MPRs, ignoring Eddington bias, while the blue lines show \citetalias{Evrard:2014} model expectations, equation~(\ref{eq:muGivenSa}), that include the mass function convolution at second order.  The lower sub-panels show the bias in the estimated halo mass, with dashed black lines showing $\pm 1\%$ accuracy with respect to the LLR true estimate. }
\label{fig:E14-test}
\end{figure}

The model estimate can be compared to the true log-mean halo mass in the simulations.  To determine the underlying ``true'' values of $\langle \mu | s_a, z \rangle$, we perform the inverse LLR fit to that used above, meaning we fit for the mean total halo mass, $M_{500}$, as a function of either stellar mass or gas mass. We perform this regression above $\mstar = 10^{12} \msol$ and $\mgas =4 \times 10^{11} \msol$.   The results are shown as black lines in the upper panels of Figure~\ref{fig:E14-test}. 

The lower panels of Figure~\ref{fig:E14-test} show the accuracy of various estimates compared to the direct LLR fits. Green lines show the naive estimator, $\langle \mu | s_a, z \rangle = (s_a - \pi_a)/ \alpha_a$, using best fit with constant slopes over halos with total masses $> 10^{13} \msun$.  This naive estimator, which ignores both the mass dependence of the slope and the Eddington bias, struggles to achieve mass accuracy at the level of $10 \%$.  

Red dashed lines improve on this naive estimate by using the local slope from the LLR model, Figure~\ref{fig:mass-evolution}, while still ignoring the Eddington correction.  This model is an improvement but it does not reach percent-level mass accuracy, given by the horizontal dotted lines in the lower panels of Figure~\ref{fig:E14-test}. 

Applying the full expression of equation~(\ref {eq:muGivenSa}), with the bias term and local estimates of the slope and scatter, leads to the blue line in Figure~\ref{fig:E14-test}. This estimate recovers the true mean mass within $1\%$ for selection by $\mgas$ over the entire mass range shown.  

Equation~(\ref {eq:muGivenSa}) is similarly accurate for selection by $\mstar$ above a stellar mass of $10^{12.3} \msun$.  Below this the error grows, approaching a 5\% bias at the lowest stellar masses.  In halos near $10^{13} \msun$ that host poor groups of galaxies, the scatter in cumulative stellar mass within halos is large, $\sigma \simeq 0.3$.  The equivalent mass scatter at fixed $\mstar$, given by $\sigma_\mu = \sigma / \alpha$ is larger, $\sigma_\mu \simeq 0.4$, since the LLR slope is sub-linear, $\alpha \sim 0.8$.  
The magnitude of the bias correction, proportional to the MPR variance, is largest for the low-mass halos selected by $\mstar$. In addition, there may be some non-Guassianity beginning to appear in $p(\mstar \ \mhalo)$ at these low masses, as close inspection of Figure~\ref{fig:E14-test} indicates.

What we have shown is that simple properties of simulated halos, namely $\mgas$ and $\mstar$, follow the \citetalias{Evrard:2014} model form at a level sufficient to achieve sub-percent accuracy in estimated log-mean total halo mass.  The test here, involving intrinsic \emph{halo} properties, $\mathbf{S}^{\rm int}$, measured directly within the simulations, is a prelude to more realistic tests using mock observables.  Projection and telescope/instrument effects introduce an extra convolution, $p(\mathbf{S}^{\rm obs} | \mathbf{S}^{\rm int},z)$, that may introduce non-Gaussianity into the form of the measured observables, $\mathbf{S}^{\rm obs}$. We defer such survey and instrument-specific studies to future work. 

Future work will extend this analysis to include additional observable properties such as X-ray temperature or luminosity.  Support for cosmological analysis also requires mapping intrinsic to observed properties in a survey-specific manner, a process that could induce non-Gaussian features into the conditional statistics.

\section{Discussion} \label{sec:discussion}

Here we discuss our findings in the context of previous simulation work.  We offer some initial thoughts on observations, but leave detailed study of modeling observed MPRs to future work.  

\subsection{Mean MPR behavior}

The cosmo-OWLS simulations, precursor to those used here, display hot gas scaling trends similar to those of \bahamas and \macsis simulations. \citet{LeBrun:2016} fit the median behavior in mass bins for halos above $10^{13} \msol$ and $0 < z  <1.5$ to both single and broken power law forms.  For $\Delta = 500$ they find a single power-law slope in $\mgas$ of $1.32 \pm 0.02$, intermediate to the values shown in Figure~\ref{fig:mass-evolution}.  Using a break point of $M_{500} = 10^{14} \msol$, they find a high-mass slope of $1.18 \pm 0.02$, similar to our LLR values at $3\times 10^{14} \msol$.  For low masses between the break and sample limit, they find redshift-dependent behavior with a slope of $1.74$ at $z=0$ declining to $1.32$ at $z=1$. The \bahamas and \macsis samples behave similarly; the local LLR slope of the $\mgas$ MPR is most sensitive to redshift below $10^{14} \msol$. 

Using an independent smoothed particle hydrodynamics code, \citet{Truong:2016} simulate 24 massive halos with astrophysical treatment that includes AGN feedback.  While their methods are not directly calibrated to match the observed gas content of clusters, their estimate of the $\mgas$ MPR slope is $\sim 1.07$, near the value found for halo masses $3 \times 10^{14} \msun$ in the \bahamas and \macsis simulations. 

The IllustrisTNG project \citep{Springel:2017} produces full-physics simulations of 100 and 300 Mpc volumes with a moving-mesh code and an updated feedback model. \citet{Pillepich:2017} study the stellar contents of a subset of halos at redshift $z<1$ derived from the TNG100 and TNG300 simulations.  Fitting a single power-law to the total stellar mass MPR around a mass scale of $M_{500}=10^{14} \msol$, they find a slope of $0.84$, in very good agreement with our findings.

The trend toward a self-similar slope of one in the $\mgas$ MPR is supported by the observational sample of relaxed, high mass clusters by \citet{Mantz:2016}.  Using weak lensing masses, they find a slope of $1.04 \pm 0.05$ in the $\mgas - M_{WL}$ relation for 40 clusters with ${\rm kT} > 5 ~{\rm keV}$.  Studies of lower mass clusters typically find super-linear scaling of gas mass with halo mass, such as the slope of $1.22 \pm 0.04$ found by \citet{Lovisari:2015} for a sample of 82 clusters.

\subsection{Diagonal elements of the property covariance}

The intrinsic scatter in the MPR for a certain property sets its quality as a proxy for total halos mass. Among observable X-ray properties, it has previously been noted that $\mgas$ has low scatter in both observations \citep{Okabe:2010,Mantz:2016} and hydrodynamic simulations 
\citep{Stanek:2010,Truong:2016,LeBrun:2016,Barnes:2017}.

For cosmo-OWLs, \citet{LeBrun:2016} find a scatter of $0.11$ in $\mgas$ at fixed halo mass of $10^{14} \msol$ at $z=0$, which agrees well with our results. They find redshift and mass trends similar to those found here.  \citet{Wu:2015} find $\mgas$ scatter of $0.08$ in the Rhapsody-G simulations of ten massive halos, including their progenitors.   \citet{Truong:2016} find a somewhat smaller scatter of $0.06$ in their sample of 24 halos.  

We note that the scatter derived in this work is an intrinsic halo property whereas the observational data are measured in a projected space.  Given the incoherent nature of projections, the scatter derived from observational data should be larger that the intrinsic values derived in this work. For instance, \citet{Mantz:2016} find $0.09 \pm 0.02$ for $\mgas$ for halos above $3 \times 10^{14} \msun$ which is marginally larger than what is found in this work.

On the scatter in overall stellar mass at fixed halo mass, relatively little work has been published from either simulations or observations.   \citet{Pillepich:2017} find scatter of $0.16$ in $\mstar$ the TNG100 and TNG300 simulations for halos $\sim 10^{14} \msol$, in good agreement with the  \bahamas and \macsis results. A more detailed comparison is needed to compare trends with mass and redshift more precisely. In the Rhapsody-G sample, \citet{Wu:2015} find $\mstar$ a larger scatter of $0.34$ in a combined sample comprised of ten massive halos at $z=0$ and their progenitors at $z=0.5$ and 1.

Observationally, \citet{Zu:2015} combine the galaxy stellar mass function with galaxy-galaxy lensing and galaxy clustering from a sample of Sloan Digital Sky Survey (SDSS) clusters and find a scatter in the natural log of central galaxy stellar mass of $0.4$ for clusters with masses near $10^{14} \msol$.  They also find statistically significant evidence in favor of the scatter in $\mstar$ decreasing with increasing halo mass, but this refers only to the central galaxy, not the total stellar content.

\subsection{The off-diagonal element of the property covariance}

In contrast to the diagonal elements which determine the mass proxy quality of individual properties, the off-diagonal covariance elements of the joint property matrix have received far less attention.  

The results presented in \S\ref{sec:covar} are from hydrodynamics simulations that have been carefully calibrated to reproduce the observed mean relations between gas mass and halo mass and stellar mass and halo mass.  While model-dependent, these theoretical predictions are testable empirically with current and future multi-wavelength survey data.

The Rhapsody-G simulation by \citet{Wu:2015} established the first estimate of anti-correlation between stellar and gaseous content of halos. In this work, we extend their analysis by using a much larger halo sample that extends to galaxy group scales. 

In agreement with \citet{Wu:2015}, we find that the most massive systems are approximately ``closed boxes'', but our correlation coefficient peaks at a smaller magnitude than the value of $-0.68$ found in that work.  For the group size halos, the link between the stellar mass and hot gas mass is strongly reduced (see Figure \ref{fig:corr}).  This trend is due to more efficient feedback in low mass halos that ejects a significant fraction of the gas from the progenitors of the groups to radii outside $R_{500}$.

Furthermore, we see redshift evolution in the correlation coefficient toward larger anti-correlation at later times.  While this finding might suggest that halos of fixed mass vent their baryonic content more efficiently at high redshift, this scenario is not supported by the LLR normalizations (Figure~\ref{fig:fgas-star}) which indicate that baryon fractions {\it increase} with increasing redshift at fixed halo mass.  However, we observe increasing scatter at lower redshift for both gas mass and stellar mass at fixed halo mass, which allows more a longer lever arm to support correlation.
Accretion events might be the key in understanding this trend.  Massive halos gain mass through merging and accretion, and the rate of accretion declines with redshift \citep{Fakhouri:2010}. Due to the stochastic nature of these events, these events add additional ``irreducible scatter'' which could weaken the strength of anti-correlation.

A key difference between the Rhapsody-G simulation results of \citet{Wu:2015} and ours is the scatter in $\mstar$ at fixed halo mass, which for high mass halos is much larger in Rhapsody-G $(>30\%)$ than \bahamas and \macsis simulation $(<10\%)$.  
We note that the Rhapsody-G sample combines all halos progenitors into a single sample. The different sample definitions, along with different numerical and modeling treatments for star formation and feedback, are likely both conspiring to create the difference in property correlation behavior.

The return toward zero of the correlation coefficient for high mass systems most likely has a simple origin: the very small effect of scatter in $\mstar$. Comparing Figures~\ref{fig:mass-evolution} and \ref{fig:fgas-star}, we see that a typical $10^{15} \msol$ halo at $z=0$ will have converted $10\%$ of its baryons into stars, with $75\%$ remaining in hot gas within $R_{500}$.  The {\sl fractional} deviations in these components are $0.1$ and $0.05$, respectively, meaning the contributions to the baryon fraction scatter are roughly $0.01$ for stars and $0.04$ for hot gas.  These small values leave little room for coupling deviations in gas mass with those in stellar mass.  By comparison, the contributions to the baryon fraction scatter at $10^{14} \msol$ are larger by roughly a factor of two, $0.02$ for stars and $0.07$ for hot gas.  

Put another way, we expect \emph{irreducible scatter} in the baryon content of halos when masses are defined using a simple spherical threshold.  Deviations are sourced by the basic nature of the dynamics --- collisionless for dark matter and stars but collisional for gas --- as well as edge effects introduced by the spherical filter, incluing choice of center.  A measure of this irreducible scatter can be found from the gravity-only models of \citet{Stanek:2010}, which show a fractional scatter in gas/baryon mass (there are no stars) at fixed halo mass of $0.036 \pm 0.001$.  This value is very close to the level seen in the hot gas phase of \bahamas and \macsis halos above $10^{15} \msol$.

We remind the reader that these are results from a model-dependent simulation.  These predictions await testing by future empirical studies, which will ultimately be capable of constraining the baryon content covariance of clusters with high accuracy.

\subsection{Observational prospects for stellar-hot gas mass covariance}

The historical absence of well-defined, uniform, multi-wavelength cluster samples explains the sparsity of observational attempts to constrain the off-diagonal elements of the property covariance matrix.  The few extant studies focus on covariance between X-ray observables \citep[e.g.][]{Mantz:2010,Maughan:2014,Mantz:2016,Andreon:2017}. To the best of our knowledge, no constraint on the correlation between an optical and X-ray property pair has been reported.  Finally, modeling the mapping between cluster observables and  intrinsic halo properties is an important task.

A minimum requirement is to obtain both stellar mass and gas mass estimates for a large cluster sample with a well-defined selection function.  
Uniformity of the sample is a key factor; combining several heterogeneous datasets is not an option due to complexity in modeling the full selection function.

The Local Cluster Substructure Survey (LoCuSS, PI:
G.P. Smith) survey\footnote{\url{http://www.sr.bham.ac.uk/locuss/}} is taking the lead to make such a measurement possible by combining multi-wavelength observables for a well-defined cluster sample of moderate size. LoCuSS will help to have a preliminary result on the value of the correlation coefficient; however, further studies with larger sample size and broader mass and redshift ranges are needed to study these quantities in more depth.

\subsection{Sensitivity to Cosmological Parameters}

To test whether our findings are sensitive to the underlying cosmology, we analyzed the \emph{WMAP9} cosmology suite of the \bahamas simulation at $z=0$, $0.5$, and $1.0$. 
We obtain results in good agreement with results from the \emph{Planck} cosmology.
Specifically, we find evidence for a log-normal PDF and see trends in LLR scaling parameters, including off diagonal elements, similar to those we report here.  This reaffirms that the log-normal assumption is a sufficient statistical model independent of cosmological parameters.

\section{Conclusion}
\label{sec:conclusion}

We present population statistics for volume-limited samples of massive halos selected from the \bahamas simulation and its high-mass extension, \macsis. The combination of these two sets of simulations provides large sample sizes across a wide dynamic range in halo mass realized with consistent, sub-grid physics treatments for star formation and feedback from supernovae and active galactic nuclei. 
We introduce local linear regression to measure conditional statistical properties of stellar mass and hot gas mass given total halo mass, including their covariance.  We assess the validity of the log-normal assumption in MPR models, and investigate the accuracy of the multi-property analytical model of \citetalias{Evrard:2014}.

Our main findings are as follows.

\begin{itemize}
\item  The scalings of $\langle \ln \mgas | \mhalo, z \rangle $ and $\langle \ln \mstar | \mhalo, z \rangle $ with halo mass are well approximated by power laws with running exponents.  For clusters with masses above $10^{14} \msol$, the local slope and scatter behave monotonically with mass.  The local slope and scatter in stellar mass are nearly redshift independent, while the hot gas slope and scatter tend to increase with increasing redshift.  Above  $5 \times 10^{14} \msol$, the behavior approaches simple self-similarity, with slopes approaching one and very small fractional scatter in baryon component masses: $0.04$ in hot gas and $0.08$ in stellar mass.  The component fractional scatter in galaxy groups near $\sim 3 \times 10^{13} \msol$ is significantly larger: $0.2$ in hot gas and $0.3$ in stellar mass.  

\item The PDF of residuals in gas and stellar mass about the local regression fit is very close to log-normal.  The deviations from normality in the intrinsic halo population are too small to bias cosmological constraints from cluster counts, but further modeling of sample selection effects and of how intrinsic properties map to those observed remains to be done. 

\item  Studying the hot gas and stellar property covariance, we find that massive halos display anti-correlation indicative of a ``Closed Box'' nature, with the box closing increasingly tighter at later times. The correlation coefficient is suppressed in lower mass halos, which are capable of venting a significant fraction of their baryons outside their virial regions, as well as in the highest mass halos, where small deviations about a small mean contribution in stellar mass has little effect on the overall baryon content of these systems.  

\item We verify that the model proposed by \citetalias{Evrard:2014} can predict the expected log total mass of property-selected halo samples with sub-percent accuracy when local MPR scaling parameters are used.
\end{itemize}

These theoretical predictions need to be confirmed or falsified through empirical evidence from analysis of observational data.  
 Future campaigns of multi-wavelength observational studies, such as XXL \citep{Pierre:2016} and DES \citep{DES-DE+:2016}, have the opportunity to test these predictions and enrich our knowledge of baryon component physics.

\section*{Acknowledgment}
We acknowledge support from NASA Chandra grant CXC-17800360.  
DJB and STK acknowledge support from STFC through grant ST/L000768/1. We thank Joop Schaye for his contributions to the \bahamas and \macsis simulations, and Elena Rasia useful discussions.
This work used the DiRAC Data Centric system at Durham University, operated by the Institute for Computational Cosmology on behalf of the STFC DiRAC HPC Facility (www.dirac.ac.uk). This equipment was funded by BIS National E-infrastructure capital grant ST/K00042X/1, STFC capital grants ST/H008519/1 and ST/K00087X/1, STFC DiRAC Operations grant ST/K003267/1 and Durham University. DiRAC is part of the National E-Infrastructure.

\bibliographystyle{mnras}
\bibliography{my_bib}

\appendix

\section{Data} \label{app:raw-data}

Figures~\ref{fig:mass-evolution} and \ref{fig:fgas-star} illustrate the mass and redshift dependence of the LLR slope, scatter and normalization at $\Delta = 500$.  In Table \ref{tab:raw-data-gas-500}, \ref{tab:raw-data-gas-200}, \ref{tab:raw-data-star-500}, and \ref{tab:raw-data-star-200}, we provide the resultant fit parameters of gas mass and stellar mass for $\Delta = 500$ and $200$\footnote{The LLR fit parameters with three significant digits will be available in the electronic version.}.

\begin{table*}
\centering
\caption{The LLR fit parameters for $\mgas-\mhalo$ relation at redshift $z=0,0.5,1$ for overdensity $\Delta=500$. For convenience, we use decimal logarithms for both the independent halo mass variable $\mu_{10} = \log_{10}({\rm M}_\Delta/ \msun)$, as well as the normalization, $\pi_{10}=\log_{10}(\mgas/ \msun)$. Also given are the local slope, $\alpha$, and scatter in the natural logarithm, $\sigma$, the diagonal component of equation (\ref{eq:r-estimator}). The error on the normalization is $<0.01$ in $\log_{10}$ basis. The quoted errors have two significant digits, and 0.00 value means that the uncertainty is $<0.01$. The LLR fit parameters with three significant digits will be available in the electronic version.}   
\label{tab:raw-data-gas-500}
\begin{tabular}{|*{10}{l|}}
\hline 
\multirow{2}{*}{$\mu_{10}$} 
                       & \multicolumn{3}{c|}{$z=0$} & \multicolumn{3}{c|}{$z=0.5$} & \multicolumn{3}{c|}{$z=1$} \\ \cline{2-10} 
                       & $\pi_{10}$ & $\alpha$ & $\sigma$ & $\pi_{10}$ & $\alpha$ & $\sigma$ & $\pi_{10}$ & $\alpha$ & $\sigma$   \\ \hline \hline
13.0  & $11.276$ & $1.54 \pm 0.01$ & $0.29 \pm 0.00$  & $11.500$ & $1.39 \pm 0.01$ & $0.22 \pm 0.00$  & $11.641$ & $1.30 \pm 0.01$ & $0.20 \pm 0.00$  \\
13.1  & $11.429$ & $1.56 \pm 0.01$ & $0.28 \pm 0.00$  & $11.639$ & $1.41 \pm 0.01$ & $0.22 \pm 0.00$  & $11.771$ & $1.32 \pm 0.01$ & $0.20 \pm 0.00$  \\
13.2  & $11.585$ & $1.58 \pm 0.01$ & $0.27 \pm 0.00$  & $11.780$ & $1.43 \pm 0.00$ & $0.21 \pm 0.00$  & $11.903$ & $1.33 \pm 0.01$ & $0.19 \pm 0.00$  \\
13.3  & $11.745$ & $1.60 \pm 0.00$ & $0.26 \pm 0.00$  & $11.924$ & $1.44 \pm 0.00$ & $0.20 \pm 0.00$  & $12.036$ & $1.34 \pm 0.01$ & $0.19 \pm 0.01$  \\
13.4  & $11.907$ & $1.62 \pm 0.00$ & $0.24 \pm 0.00$  & $12.069$ & $1.45 \pm 0.00$ & $0.19 \pm 0.00$  & $12.171$ & $1.34 \pm 0.01$ & $0.20 \pm 0.01$  \\
13.5  & $12.070$ & $1.63 \pm 0.00$ & $0.23 \pm 0.00$  & $12.216$ & $1.46 \pm 0.00$ & $0.18 \pm 0.00$  & $12.305$ & $1.34 \pm 0.01$ & $0.21 \pm 0.02$  \\
13.6  & $12.233$ & $1.63 \pm 0.00$ & $0.22 \pm 0.00$  & $12.362$ & $1.46 \pm 0.00$ & $0.17 \pm 0.00$  & $12.440$ & $1.34 \pm 0.01$ & $0.22 \pm 0.03$  \\
13.7  & $12.395$ & $1.62 \pm 0.00$ & $0.20 \pm 0.00$  & $12.506$ & $1.45 \pm 0.00$ & $0.16 \pm 0.00$  & $12.573$ & $1.34 \pm 0.01$ & $0.24 \pm 0.04$  \\
13.8  & $12.553$ & $1.59 \pm 0.00$ & $0.19 \pm 0.00$  & $12.647$ & $1.43 \pm 0.00$ & $0.16 \pm 0.01$  & $12.704$ & $1.32 \pm 0.01$ & $0.25 \pm 0.04$  \\
13.9  & $12.706$ & $1.55 \pm 0.00$ & $0.18 \pm 0.00$  & $12.785$ & $1.40 \pm 0.01$ & $0.17 \pm 0.01$  & $12.831$ & $1.30 \pm 0.01$ & $0.26 \pm 0.03$  \\
14.0  & $12.854$ & $1.51 \pm 0.00$ & $0.17 \pm 0.00$  & $12.917$ & $1.36 \pm 0.01$ & $0.20 \pm 0.02$  & $12.956$ & $1.27 \pm 0.01$ & $0.27 \pm 0.04$  \\
14.1  & $12.996$ & $1.45 \pm 0.01$ & $0.17 \pm 0.01$  & $13.045$ & $1.31 \pm 0.01$ & $0.23 \pm 0.03$  & $13.079$ & $1.25 \pm 0.01$ & $0.27 \pm 0.04$  \\
14.2  & $13.131$ & $1.39 \pm 0.01$ & $0.18 \pm 0.01$  & $13.168$ & $1.27 \pm 0.01$ & $0.26 \pm 0.03$  & $13.202$ & $1.24 \pm 0.01$ & $0.27 \pm 0.04$  \\
14.3  & $13.258$ & $1.32 \pm 0.01$ & $0.20 \pm 0.02$  & $13.290$ & $1.24 \pm 0.01$ & $0.28 \pm 0.04$  & $13.324$ & $1.23 \pm 0.01$ & $0.24 \pm 0.04$  \\
14.4  & $13.378$ & $1.25 \pm 0.01$ & $0.23 \pm 0.03$  & $13.411$ & $1.23 \pm 0.01$ & $0.28 \pm 0.04$  & $13.443$ & $1.20 \pm 0.01$ & $0.19 \pm 0.03$  \\
14.5  & $13.496$ & $1.21 \pm 0.01$ & $0.26 \pm 0.03$  & $13.532$ & $1.22 \pm 0.01$ & $0.25 \pm 0.04$  & $13.559$ & $1.17 \pm 0.01$ & $0.14 \pm 0.02$  \\
14.6  & $13.614$ & $1.19 \pm 0.01$ & $0.27 \pm 0.04$  & $13.652$ & $1.19 \pm 0.01$ & $0.20 \pm 0.03$  & $13.672$ & $1.14 \pm 0.01$ & $0.10 \pm 0.02$  \\
14.7  & $13.732$ & $1.18 \pm 0.01$ & $0.26 \pm 0.04$  & $13.767$ & $1.16 \pm 0.01$ & $0.14 \pm 0.02$  & $13.781$ & $1.11 \pm 0.01$ & $0.07 \pm 0.01$  \\
14.8  & $13.849$ & $1.16 \pm 0.02$ & $0.23 \pm 0.04$  & $13.878$ & $1.12 \pm 0.01$ & $0.10 \pm 0.01$  & $13.888$ & $1.09 \pm 0.01$ & $0.06 \pm 0.01$  \\
14.9  & $13.962$ & $1.13 \pm 0.02$ & $0.19 \pm 0.03$  & $13.986$ & $1.08 \pm 0.01$ & $0.07 \pm 0.01$  & $13.993$ & $1.07 \pm 0.01$ & $0.05 \pm 0.00$  \\
15.0  & $14.072$ & $1.09 \pm 0.02$ & $0.16 \pm 0.04$  & $14.091$ & $1.06 \pm 0.01$ & $0.05 \pm 0.00$  & $14.094$ & $1.05 \pm 0.01$ & $0.04 \pm 0.00$  \\
15.1  & $14.179$ & $1.06 \pm 0.02$ & $0.14 \pm 0.06$  & $14.195$ & $1.05 \pm 0.01$ & $0.04 \pm 0.00$  & $14.193$ & $1.03 \pm 0.02$ & $0.04 \pm 0.00$  \\
15.2  & $14.283$ & $1.05 \pm 0.01$ & $0.14 \pm 0.07$  & $14.297$ & $1.04 \pm 0.01$ & $0.04 \pm 0.00$  & $14.290$ & $1.01 \pm 0.02$ & $0.04 \pm 0.00$  \\
15.3  & $14.388$ & $1.05 \pm 0.01$ & $0.14 \pm 0.08$  & $14.400$ & $1.03 \pm 0.01$ & $0.04 \pm 0.00$  & $14.387$ & $1.00 \pm 0.02$ & $0.03 \pm 0.01$  \\
15.4  & $14.494$ & $1.05 \pm 0.02$ & $0.12 \pm 0.07$  & $14.503$ & $1.03 \pm 0.01$ & $0.04 \pm 0.00$  & $14.483$ & $0.99 \pm 0.02$ & $0.02 \pm 0.01$  \\
\hline
\end{tabular}
\end{table*}

\begin{table*}
\centering
\caption{The LLR fit parameters for $\mgas-\mhalo$ relation at redshift $z=0,0.5,1$ for overdensity $\Delta=200$. For convenience, we use decimal logarithms for both the independent halo mass variable $\mu_{10} = \log_{10}({\rm M}_\Delta/ \msun)$, as well as the normalization, $\pi_{10}=\log_{10}(\mgas/ \msun)$. Also given are the local slope, $\alpha$, and scatter in the natural logarithm, $\sigma$, the diagonal component of equation (\ref{eq:r-estimator}). The error on the normalization is $<0.01$ in $\log_{10}$ basis. The quoted errors have two significant digits, and 0.00 value means that the uncertainty is $<0.01$. The LLR fit parameters with three significant digits will be available in the electronic version.}
\label{tab:raw-data-gas-200}
\begin{tabular}{|*{10}{c|}}
\hline 
\multirow{2}{*}{$\mu_{10}$} 
                       & \multicolumn{3}{c|}{$z=0$} & \multicolumn{3}{c|}{$z=0.5$} & \multicolumn{3}{c|}{$z=1$} \\ \cline{2-10} 
                       & $\pi_{10}$ & $\alpha$ & $\sigma$ & $\pi_{10}$ & $\alpha$ & $\sigma$ & $\pi_{10}$ & $\alpha$ & $\sigma$   \\ \hline \hline
13.0  & $11.365$ & $1.56 \pm 0.01$ & $0.25 \pm 0.00$  & $11.574$ & $1.42 \pm 0.01$ & $0.19 \pm 0.00$  & $11.701$ & $1.32 \pm 0.01$ & $0.17 \pm 0.00$  \\
13.1  & $11.520$ & $1.56 \pm 0.01$ & $0.25 \pm 0.00$  & $11.716$ & $1.42 \pm 0.01$ & $0.19 \pm 0.00$  & $11.833$ & $1.32 \pm 0.01$ & $0.17 \pm 0.00$  \\
13.2  & $11.676$ & $1.57 \pm 0.01$ & $0.24 \pm 0.00$  & $11.857$ & $1.42 \pm 0.01$ & $0.18 \pm 0.00$  & $11.965$ & $1.32 \pm 0.01$ & $0.16 \pm 0.00$  \\
13.3  & $11.833$ & $1.57 \pm 0.01$ & $0.23 \pm 0.00$  & $11.999$ & $1.42 \pm 0.00$ & $0.18 \pm 0.00$  & $12.097$ & $1.32 \pm 0.00$ & $0.16 \pm 0.00$  \\
13.4  & $11.990$ & $1.58 \pm 0.00$ & $0.22 \pm 0.00$  & $12.140$ & $1.42 \pm 0.00$ & $0.17 \pm 0.00$  & $12.229$ & $1.32 \pm 0.00$ & $0.15 \pm 0.00$  \\
13.5  & $12.147$ & $1.57 \pm 0.00$ & $0.21 \pm 0.00$  & $12.282$ & $1.41 \pm 0.00$ & $0.16 \pm 0.00$  & $12.360$ & $1.31 \pm 0.00$ & $0.16 \pm 0.01$  \\
13.6  & $12.304$ & $1.56 \pm 0.00$ & $0.20 \pm 0.00$  & $12.422$ & $1.40 \pm 0.00$ & $0.15 \pm 0.00$  & $12.490$ & $1.30 \pm 0.00$ & $0.16 \pm 0.01$  \\
13.7  & $12.458$ & $1.55 \pm 0.00$ & $0.19 \pm 0.00$  & $12.560$ & $1.39 \pm 0.00$ & $0.14 \pm 0.00$  & $12.619$ & $1.29 \pm 0.01$ & $0.17 \pm 0.02$  \\
13.8  & $12.608$ & $1.52 \pm 0.00$ & $0.18 \pm 0.00$  & $12.696$ & $1.37 \pm 0.00$ & $0.13 \pm 0.00$  & $12.746$ & $1.28 \pm 0.01$ & $0.19 \pm 0.03$  \\
13.9  & $12.755$ & $1.48 \pm 0.00$ & $0.17 \pm 0.00$  & $12.828$ & $1.34 \pm 0.00$ & $0.13 \pm 0.00$  & $12.870$ & $1.26 \pm 0.01$ & $0.20 \pm 0.03$  \\
14.0  & $12.897$ & $1.44 \pm 0.00$ & $0.15 \pm 0.00$  & $12.958$ & $1.31 \pm 0.00$ & $0.13 \pm 0.01$  & $12.991$ & $1.23 \pm 0.01$ & $0.21 \pm 0.03$  \\
14.1  & $13.033$ & $1.39 \pm 0.00$ & $0.14 \pm 0.00$  & $13.083$ & $1.28 \pm 0.01$ & $0.14 \pm 0.01$  & $13.108$ & $1.20 \pm 0.01$ & $0.22 \pm 0.03$  \\
14.2  & $13.164$ & $1.34 \pm 0.00$ & $0.13 \pm 0.00$  & $13.203$ & $1.24 \pm 0.01$ & $0.16 \pm 0.02$  & $13.224$ & $1.18 \pm 0.01$ & $0.23 \pm 0.03$  \\
14.3  & $13.290$ & $1.29 \pm 0.00$ & $0.13 \pm 0.01$  & $13.319$ & $1.19 \pm 0.01$ & $0.19 \pm 0.02$  & $13.340$ & $1.16 \pm 0.01$ & $0.24 \pm 0.04$  \\
14.4  & $13.410$ & $1.24 \pm 0.01$ & $0.14 \pm 0.01$  & $13.432$ & $1.16 \pm 0.01$ & $0.22 \pm 0.03$  & $13.456$ & $1.16 \pm 0.01$ & $0.23 \pm 0.04$  \\
14.5  & $13.525$ & $1.18 \pm 0.01$ & $0.16 \pm 0.02$  & $13.544$ & $1.14 \pm 0.01$ & $0.24 \pm 0.03$  & $13.570$ & $1.15 \pm 0.01$ & $0.20 \pm 0.04$  \\
14.6  & $13.635$ & $1.14 \pm 0.01$ & $0.20 \pm 0.03$  & $13.658$ & $1.14 \pm 0.01$ & $0.24 \pm 0.03$  & $13.682$ & $1.13 \pm 0.01$ & $0.16 \pm 0.03$  \\
14.7  & $13.745$ & $1.11 \pm 0.01$ & $0.22 \pm 0.03$  & $13.772$ & $1.14 \pm 0.01$ & $0.21 \pm 0.03$  & $13.792$ & $1.10 \pm 0.01$ & $0.12 \pm 0.02$  \\
14.8  & $13.855$ & $1.11 \pm 0.01$ & $0.23 \pm 0.03$  & $13.884$ & $1.12 \pm 0.01$ & $0.17 \pm 0.03$  & $13.898$ & $1.08 \pm 0.01$ & $0.09 \pm 0.02$  \\
14.9  & $13.967$ & $1.11 \pm 0.01$ & $0.22 \pm 0.03$  & $13.993$ & $1.09 \pm 0.01$ & $0.12 \pm 0.02$  & $14.003$ & $1.06 \pm 0.01$ & $0.06 \pm 0.01$  \\
15.0  & $14.077$ & $1.10 \pm 0.01$ & $0.19 \pm 0.03$  & $14.098$ & $1.06 \pm 0.01$ & $0.08 \pm 0.01$  & $14.107$ & $1.05 \pm 0.01$ & $0.05 \pm 0.01$  \\
15.1  & $14.186$ & $1.08 \pm 0.01$ & $0.15 \pm 0.03$  & $14.202$ & $1.04 \pm 0.01$ & $0.06 \pm 0.01$  & $14.210$ & $1.04 \pm 0.01$ & $0.04 \pm 0.01$  \\
15.2  & $14.292$ & $1.06 \pm 0.01$ & $0.12 \pm 0.03$  & $14.303$ & $1.02 \pm 0.01$ & $0.04 \pm 0.00$  & $14.313$ & $1.04 \pm 0.01$ & $0.03 \pm 0.00$  \\
15.3  & $14.396$ & $1.04 \pm 0.01$ & $0.11 \pm 0.04$  & $14.404$ & $1.02 \pm 0.01$ & $0.03 \pm 0.00$  & $14.415$ & $1.03 \pm 0.01$ & $0.03 \pm 0.00$  \\
15.4  & $14.499$ & $1.03 \pm 0.01$ & $0.09 \pm 0.04$  & $14.505$ & $1.01 \pm 0.01$ & $0.03 \pm 0.00$  & $14.517$ & $1.03 \pm 0.01$ & $0.02 \pm 0.00$  \\
\hline
\end{tabular}
\end{table*}

\begin{table*}
\centering
\caption{The LLR fit parameters for $\mstar-\mhalo$ relation at redshift $z=0,0.5,1$ for overdensity $\Delta=500$. For convenience, we use decimal logarithms for both the independent halo mass variable $\mu_{10} = \log_{10}({\rm M}_\Delta/ \msun)$, as well as the normalization, $\pi_{10}=\log_{10}(\mgas/ \msun)$. Also given are the local slope, $\alpha$, and scatter in the natural logarithm, $\sigma$, the diagonal component of equation (\ref{eq:r-estimator}). The error on the normalization is $<0.01$ in $\log_{10}$ basis. The quoted errors have two significant digits, and 0.00 value means that the uncertainty is $<0.01$. The LLR fit parameters with three significant digits will be available in the electronic version.}
\label{tab:raw-data-star-500}
\begin{tabular}{|*{10}{c|}}
\hline 
\multirow{2}{*}{$\mu_{10}$} 
                       & \multicolumn{3}{c|}{$z=0$} & \multicolumn{3}{c|}{$z=0.5$} & \multicolumn{3}{c|}{$z=1$} \\ \cline{2-10} 
                       & $\pi_{10}$ & $\alpha$ & $\sigma$ & $\pi_{10}$ & $\alpha$ & $\sigma$ & $\pi_{10}$ & $\alpha$ & $\sigma$   \\ \hline \hline
13.0  & $11.432$ & $0.78 \pm 0.01$ & $0.37 \pm 0.00$  & $11.376$ & $0.77 \pm 0.01$ & $0.34 \pm 0.00$  & $11.328$ & $0.79 \pm 0.01$ & $0.32 \pm 0.00$  \\
13.1  & $11.510$ & $0.78 \pm 0.01$ & $0.36 \pm 0.00$  & $11.453$ & $0.78 \pm 0.01$ & $0.34 \pm 0.00$  & $11.407$ & $0.80 \pm 0.01$ & $0.31 \pm 0.00$  \\
13.2  & $11.588$ & $0.79 \pm 0.01$ & $0.35 \pm 0.00$  & $11.531$ & $0.78 \pm 0.01$ & $0.33 \pm 0.00$  & $11.487$ & $0.81 \pm 0.01$ & $0.30 \pm 0.00$  \\
13.3  & $11.667$ & $0.79 \pm 0.01$ & $0.33 \pm 0.00$  & $11.609$ & $0.79 \pm 0.01$ & $0.31 \pm 0.00$  & $11.569$ & $0.82 \pm 0.01$ & $0.29 \pm 0.00$  \\
13.4  & $11.747$ & $0.79 \pm 0.00$ & $0.32 \pm 0.00$  & $11.690$ & $0.80 \pm 0.00$ & $0.30 \pm 0.00$  & $11.653$ & $0.83 \pm 0.01$ & $0.28 \pm 0.00$  \\
13.5  & $11.827$ & $0.80 \pm 0.00$ & $0.30 \pm 0.00$  & $11.772$ & $0.81 \pm 0.00$ & $0.28 \pm 0.00$  & $11.737$ & $0.84 \pm 0.01$ & $0.26 \pm 0.00$  \\
13.6  & $11.907$ & $0.80 \pm 0.00$ & $0.28 \pm 0.00$  & $11.855$ & $0.83 \pm 0.00$ & $0.26 \pm 0.00$  & $11.823$ & $0.85 \pm 0.01$ & $0.25 \pm 0.00$  \\
13.7  & $11.988$ & $0.81 \pm 0.00$ & $0.25 \pm 0.00$  & $11.939$ & $0.84 \pm 0.00$ & $0.24 \pm 0.00$  & $11.909$ & $0.86 \pm 0.01$ & $0.23 \pm 0.00$  \\
13.8  & $12.071$ & $0.82 \pm 0.00$ & $0.23 \pm 0.00$  & $12.025$ & $0.85 \pm 0.00$ & $0.22 \pm 0.00$  & $11.996$ & $0.86 \pm 0.01$ & $0.22 \pm 0.01$  \\
13.9  & $12.154$ & $0.83 \pm 0.00$ & $0.22 \pm 0.00$  & $12.110$ & $0.85 \pm 0.01$ & $0.21 \pm 0.00$  & $12.081$ & $0.86 \pm 0.01$ & $0.21 \pm 0.01$  \\
14.0  & $12.238$ & $0.84 \pm 0.01$ & $0.20 \pm 0.00$  & $12.196$ & $0.86 \pm 0.01$ & $0.20 \pm 0.01$  & $12.166$ & $0.85 \pm 0.01$ & $0.21 \pm 0.01$  \\
14.1  & $12.323$ & $0.84 \pm 0.01$ & $0.19 \pm 0.00$  & $12.282$ & $0.86 \pm 0.01$ & $0.20 \pm 0.01$  & $12.252$ & $0.86 \pm 0.01$ & $0.21 \pm 0.02$  \\
14.2  & $12.408$ & $0.85 \pm 0.01$ & $0.18 \pm 0.01$  & $12.369$ & $0.87 \pm 0.01$ & $0.21 \pm 0.01$  & $12.342$ & $0.88 \pm 0.01$ & $0.20 \pm 0.02$  \\
14.3  & $12.494$ & $0.85 \pm 0.01$ & $0.18 \pm 0.01$  & $12.459$ & $0.88 \pm 0.01$ & $0.21 \pm 0.02$  & $12.434$ & $0.91 \pm 0.01$ & $0.19 \pm 0.02$  \\
14.4  & $12.581$ & $0.86 \pm 0.01$ & $0.19 \pm 0.01$  & $12.551$ & $0.91 \pm 0.01$ & $0.20 \pm 0.02$  & $12.528$ & $0.93 \pm 0.01$ & $0.16 \pm 0.02$  \\
14.5  & $12.671$ & $0.89 \pm 0.01$ & $0.20 \pm 0.02$  & $12.646$ & $0.93 \pm 0.01$ & $0.18 \pm 0.02$  & $12.623$ & $0.95 \pm 0.01$ & $0.14 \pm 0.01$  \\
14.6  & $12.763$ & $0.91 \pm 0.01$ & $0.21 \pm 0.02$  & $12.741$ & $0.94 \pm 0.01$ & $0.15 \pm 0.02$  & $12.719$ & $0.95 \pm 0.01$ & $0.12 \pm 0.01$  \\
14.7  & $12.858$ & $0.94 \pm 0.01$ & $0.19 \pm 0.02$  & $12.836$ & $0.95 \pm 0.01$ & $0.12 \pm 0.01$  & $12.815$ & $0.95 \pm 0.01$ & $0.11 \pm 0.01$  \\
14.8  & $12.954$ & $0.95 \pm 0.01$ & $0.17 \pm 0.02$  & $12.930$ & $0.95 \pm 0.01$ & $0.10 \pm 0.01$  & $12.909$ & $0.95 \pm 0.02$ & $0.10 \pm 0.01$  \\
14.9  & $13.050$ & $0.96 \pm 0.01$ & $0.14 \pm 0.01$  & $13.025$ & $0.95 \pm 0.01$ & $0.09 \pm 0.00$  & $13.003$ & $0.95 \pm 0.02$ & $0.09 \pm 0.01$  \\
15.0  & $13.145$ & $0.96 \pm 0.01$ & $0.11 \pm 0.01$  & $13.121$ & $0.95 \pm 0.01$ & $0.09 \pm 0.00$  & $13.098$ & $0.95 \pm 0.02$ & $0.08 \pm 0.01$  \\
15.1  & $13.241$ & $0.96 \pm 0.01$ & $0.09 \pm 0.00$  & $13.217$ & $0.96 \pm 0.01$ & $0.08 \pm 0.00$  & $13.195$ & $0.95 \pm 0.02$ & $0.08 \pm 0.01$  \\
15.2  & $13.337$ & $0.96 \pm 0.01$ & $0.08 \pm 0.00$  & $13.312$ & $0.96 \pm 0.02$ & $0.08 \pm 0.00$  & $13.292$ & $0.96 \pm 0.02$ & $0.07 \pm 0.01$  \\
15.3  & $13.433$ & $0.96 \pm 0.01$ & $0.08 \pm 0.00$  & $13.409$ & $0.96 \pm 0.02$ & $0.08 \pm 0.01$  & $13.389$ & $0.96 \pm 0.02$ & $0.06 \pm 0.01$  \\
15.4  & $13.529$ & $0.96 \pm 0.01$ & $0.08 \pm 0.00$  & $13.510$ & $0.98 \pm 0.03$ & $0.08 \pm 0.01$  & $13.487$ & $0.97 \pm 0.03$ & $0.04 \pm 0.01$  \\
\hline
\end{tabular}
\end{table*}

\begin{table*}
\centering
\caption{The LLR fit parameters for $\mstar-\mhalo$ relation at redshift $z=0,0.5,1$ for overdensity $\Delta=200$. For convenience, we use decimal logarithms for both the independent halo mass variable $\mu_{10} = \log_{10}({\rm M}_\Delta/ \msun)$, as well as the normalization, $\pi_{10}=\log_{10}(\mgas/ \msun)$. Also given are the local slope, $\alpha$, and scatter in the natural logarithm, $\sigma$, the diagonal component of equation (\ref{eq:r-estimator}). The error on the normalization is $<0.01$ in $\log_{10}$ basis. The quoted errors have two significant digits, and 0.00 value means that the uncertainty is $<0.01$. The LLR fit parameters with three significant digits will be available in the electronic version.}
\label{tab:raw-data-star-200}
\begin{tabular}{|*{10}{c|}}
\hline 
\multirow{2}{*}{$\mu_{10}$} 
                       & \multicolumn{3}{c|}{$z=0$} & \multicolumn{3}{c|}{$z=0.5$} & \multicolumn{3}{c|}{$z=1$} \\ \cline{2-10} 
                       & $\pi_{10}$ & $\alpha$ & $\sigma$ & $\pi_{10}$ & $\alpha$ & $\sigma$ & $\pi_{10}$ & $\alpha$ & $\sigma$   \\ \hline \hline
13.0  & $11.347$ & $0.87 \pm 0.01$ & $0.36 \pm 0.00$  & $11.298$ & $0.83 \pm 0.02$ & $0.34 \pm 0.00$  & $11.262$ & $0.77 \pm 0.02$ & $0.31 \pm 0.00$  \\
13.1  & $11.436$ & $0.86 \pm 0.01$ & $0.36 \pm 0.00$  & $11.380$ & $0.83 \pm 0.01$ & $0.33 \pm 0.00$  & $11.335$ & $0.80 \pm 0.01$ & $0.31 \pm 0.00$  \\
13.2  & $11.522$ & $0.85 \pm 0.01$ & $0.35 \pm 0.00$  & $11.463$ & $0.84 \pm 0.01$ & $0.32 \pm 0.00$  & $11.413$ & $0.82 \pm 0.01$ & $0.30 \pm 0.00$  \\
13.3  & $11.607$ & $0.84 \pm 0.01$ & $0.34 \pm 0.00$  & $11.546$ & $0.84 \pm 0.01$ & $0.31 \pm 0.00$  & $11.495$ & $0.83 \pm 0.01$ & $0.29 \pm 0.00$  \\
13.4  & $11.691$ & $0.83 \pm 0.01$ & $0.32 \pm 0.00$  & $11.630$ & $0.84 \pm 0.01$ & $0.30 \pm 0.00$  & $11.579$ & $0.85 \pm 0.01$ & $0.28 \pm 0.00$  \\
13.5  & $11.774$ & $0.83 \pm 0.00$ & $0.31 \pm 0.00$  & $11.714$ & $0.84 \pm 0.00$ & $0.29 \pm 0.00$  & $11.665$ & $0.86 \pm 0.01$ & $0.27 \pm 0.00$  \\
13.6  & $11.857$ & $0.83 \pm 0.00$ & $0.29 \pm 0.00$  & $11.799$ & $0.84 \pm 0.00$ & $0.27 \pm 0.00$  & $11.752$ & $0.87 \pm 0.01$ & $0.25 \pm 0.00$  \\
13.7  & $11.941$ & $0.83 \pm 0.00$ & $0.27 \pm 0.00$  & $11.884$ & $0.85 \pm 0.00$ & $0.25 \pm 0.00$  & $11.840$ & $0.88 \pm 0.01$ & $0.24 \pm 0.00$  \\
13.8  & $12.024$ & $0.84 \pm 0.00$ & $0.25 \pm 0.00$  & $11.970$ & $0.86 \pm 0.00$ & $0.23 \pm 0.00$  & $11.929$ & $0.88 \pm 0.01$ & $0.22 \pm 0.00$  \\
13.9  & $12.109$ & $0.84 \pm 0.00$ & $0.23 \pm 0.00$  & $12.057$ & $0.87 \pm 0.00$ & $0.21 \pm 0.00$  & $12.018$ & $0.89 \pm 0.01$ & $0.21 \pm 0.00$  \\
14.0  & $12.195$ & $0.85 \pm 0.00$ & $0.21 \pm 0.00$  & $12.145$ & $0.87 \pm 0.00$ & $0.20 \pm 0.00$  & $12.107$ & $0.89 \pm 0.01$ & $0.20 \pm 0.00$  \\
14.1  & $12.281$ & $0.86 \pm 0.00$ & $0.19 \pm 0.00$  & $12.233$ & $0.88 \pm 0.01$ & $0.19 \pm 0.00$  & $12.194$ & $0.88 \pm 0.01$ & $0.19 \pm 0.01$  \\
14.2  & $12.368$ & $0.87 \pm 0.00$ & $0.18 \pm 0.00$  & $12.321$ & $0.88 \pm 0.01$ & $0.18 \pm 0.01$  & $12.281$ & $0.87 \pm 0.01$ & $0.19 \pm 0.01$  \\
14.3  & $12.456$ & $0.87 \pm 0.01$ & $0.17 \pm 0.00$  & $12.409$ & $0.88 \pm 0.01$ & $0.19 \pm 0.01$  & $12.370$ & $0.88 \pm 0.01$ & $0.19 \pm 0.02$  \\
14.4  & $12.543$ & $0.88 \pm 0.01$ & $0.16 \pm 0.00$  & $12.498$ & $0.89 \pm 0.01$ & $0.19 \pm 0.01$  & $12.462$ & $0.91 \pm 0.01$ & $0.18 \pm 0.02$  \\
14.5  & $12.631$ & $0.88 \pm 0.01$ & $0.17 \pm 0.01$  & $12.590$ & $0.90 \pm 0.01$ & $0.20 \pm 0.02$  & $12.557$ & $0.93 \pm 0.01$ & $0.17 \pm 0.02$  \\
14.6  & $12.720$ & $0.89 \pm 0.01$ & $0.18 \pm 0.01$  & $12.684$ & $0.93 \pm 0.01$ & $0.19 \pm 0.02$  & $12.653$ & $0.96 \pm 0.01$ & $0.15 \pm 0.01$  \\
14.7  & $12.811$ & $0.90 \pm 0.01$ & $0.19 \pm 0.02$  & $12.780$ & $0.95 \pm 0.01$ & $0.17 \pm 0.02$  & $12.752$ & $0.98 \pm 0.01$ & $0.13 \pm 0.01$  \\
14.8  & $12.905$ & $0.93 \pm 0.01$ & $0.19 \pm 0.02$  & $12.877$ & $0.96 \pm 0.01$ & $0.15 \pm 0.02$  & $12.851$ & $0.99 \pm 0.01$ & $0.11 \pm 0.01$  \\
14.9  & $13.000$ & $0.95 \pm 0.01$ & $0.18 \pm 0.02$  & $12.974$ & $0.97 \pm 0.01$ & $0.12 \pm 0.01$  & $12.950$ & $0.99 \pm 0.01$ & $0.09 \pm 0.01$  \\
15.0  & $13.097$ & $0.96 \pm 0.01$ & $0.15 \pm 0.02$  & $13.071$ & $0.97 \pm 0.01$ & $0.10 \pm 0.01$  & $13.048$ & $0.98 \pm 0.01$ & $0.08 \pm 0.01$  \\
15.1  & $13.194$ & $0.97 \pm 0.01$ & $0.12 \pm 0.01$  & $13.169$ & $0.98 \pm 0.01$ & $0.08 \pm 0.00$  & $13.147$ & $0.99 \pm 0.02$ & $0.08 \pm 0.01$  \\
15.2  & $13.291$ & $0.97 \pm 0.01$ & $0.10 \pm 0.01$  & $13.268$ & $0.98 \pm 0.01$ & $0.08 \pm 0.00$  & $13.248$ & $0.99 \pm 0.02$ & $0.07 \pm 0.01$  \\
15.3  & $13.388$ & $0.97 \pm 0.01$ & $0.08 \pm 0.00$  & $13.366$ & $0.98 \pm 0.01$ & $0.07 \pm 0.00$  & $13.350$ & $1.00 \pm 0.03$ & $0.07 \pm 0.01$  \\
15.4  & $13.486$ & $0.98 \pm 0.01$ & $0.07 \pm 0.00$  & $13.465$ & $0.99 \pm 0.02$ & $0.07 \pm 0.00$  & $13.452$ & $1.01 \pm 0.04$ & $0.06 \pm 0.01$  \\
\hline
\end{tabular}
\end{table*}

\label{lastpage}

\end{document}